\documentclass[aps,pra,twocolumn,amsmath,amssymb,footinbib,showpacs,longbibliography,superscriptaddress]{revtex4-1}

\usepackage{times}
\usepackage[english]{babel}
\usepackage{latexsym}
\usepackage{graphics}
\usepackage{subfigure}
\usepackage{epsfig}
\usepackage{color}
\usepackage{hyperref}
\usepackage{braket} 
\usepackage{amstext}
\usepackage{amssymb}
\usepackage{graphicx}
\usepackage{esint}
\usepackage{braket}
\usepackage{babel}
\usepackage{amsmath}
\hypersetup{
colorlinks=true,
citecolor=blue,
linkcolor=red,
urlcolor=black
}


\begin{document}

\title{Quench Spectroscopy for Dissipative and (Non)-Hermitian Quantum Lattice Models}

\author{Julien Despres}
\affiliation{JEIP, USR 3573 CNRS, Coll\`ege de France, PSL Research University, 11 Place Marcelin Berthelot, 75321 Paris Cedex 05, France}

\date{\today}

\begin{abstract}
We extend the quench spectroscopy method to dissipative and isolated non-Hermitian quantum lattice models via the case study of the open Bose-Hubbard chain and the non-Hermitian transverse-field Ising chain respectively.
We first investigate theoretically the dynamics of the open Bose-Hubbard chain confined in the superfluid phase induced by a sudden global quench on the dissipations and the repulsive interactions using the equation-of-motion approach. Using the same analytical approach, we then discuss the applicability of the quench spectroscopy to non-Hermitian quantum lattice models by considering the sudden global quench dynamics of the non-Hermitian transverse-field Ising chain confined in the paramagnetic phase. We finally generalize this spectroscopy method to isolated Hermitian quantum lattice models characterized by a quadratic fermionic or bosonic Hamiltonian. For this purpose, we consider the case study of the Hermitian version of the latter one-dimensional lattice model. The investigation is performed analytically for the bosonic and fermionic reformulations while considering for each case the equation-of-motion and quasiparticle theoretical approaches.  
\end{abstract}

\maketitle

\section{Introduction}
\label{sec:intro}
In the last decades, the major progress in the experimental control of quantum lattice models has given consequent momentum to the analytical and experimental  
investigation of the quench dynamics of isolated Hermitian quantum systems~\cite{jurcevic2014,richerme2014,cheneau2012,polkovnikov2011,gogolin2016,calabrese2005}. Recently, a new spectroscopy method has been introduced theoretically and verified numerically using tensor-network-based techniques for isolated quantum spin and bosonic lattice models. The latter, justified using an eigenstate decomposition, is based on the dynamical properties of the model induced by weak sudden global or local quenches~\cite{sanchezpalencia2019,villa2020} and is referred to as quench
spectroscopy. This new method to determine low-lying excitation spectra has been recently applied experimentally to the dipolar XY model using neutral Rydberg atoms~\cite{browaeys2024} and has provided promising results. However, since quantum systems cannot be purely isolated from their environment, it is natural that the study of open quantum systems has attracted a lot of interest in the last few years~\cite{zoller2008,buchler2008,cirac2009} by taking into account dissipative effects such as local dephasing noise~\cite{esposito2005,gaspard2005,Eisler_2011,kollath2018,turkeshi2021diffusion}, incoherent hopping~\cite{alba2020} and local gain/loss processes ~\cite{langen2016,rauer2016,syassen2008,bo2013,ott2013,alba2022,rosso2021,rosso2023,mazza2023dissipative}. These open quantum lattice models are governed by the well-known Lindblad master equation ~\cite{fazio2024,breuer2007,ashida2020,daley2014} characterizing the time evolution of the quantum system density matrix. However, the possibility to apply the quench spectroscopy to open and non-Hermitian quantum lattice models has not been investigated and corresponds to the central point of this research work.

Experimentally, quantum simulators based on trapped ultra-cold atoms loaded in an artificial optical lattice generated by the interference of counter-propagating laser beams permit to simulate quantum lattice models~\cite{bloch2012,gross2017}. Using bosonic atoms, the Bose-Hubbard model can thus be engineered~\cite{bakr2010,chen2011}. Its out-of-equilibrium properties are accessible via quantum quenches realized experimentally by suddenly modifying the intensity of the laser beams controlling the artificial lattice depth and thus the ratio between the hopping amplitude and the two-body repulsive interaction strength~\cite{cheneau2012,trotzky2012}. This tuning of the interaction ratio offers the possibility to scan the quantum phase transition between the gapless superfluid and gapped Mott-insulating phases for commensurate fillings of the lattice. Loss processes, inducing a rich and new physics such as the quantum Zeno effect~\cite{schiro2022}, the loss-assisted quantum control~\cite{dong2022} or the loss-induced cooling effect~\cite{kohl2024}, can be generated using quantum simulators based on ultra-cold bosonic atoms where different number of particles can be involved. For instance, two-body losses can be engineered by light-assisted inelastic two-body collisions which can also occur naturally within this experimental platform~\cite{aspect2000,franchi2017,tomita2017,syassen2008,weiner1999,bouganne2020}. 

In this work, we propose to investigate the dynamics of the open Bose-Hubbard chain driven out of equilibrium via a sudden global quench on both the interaction and dissipation strengths. The dissipative effects consist here in on-site two-body losses. The term \textit{on-site} refers to a loss process involving a single lattice site. This specific quantum lattice model as well as its quench dynamics can be simulated experimentally using ultra-cold bosonic atoms~\cite{greiner2002b,cheneau2012,langen2013,geiger2014} including a controllable strength of the on-site two-body losses engineered by light-assisted inelastic collisions~\cite{tomita2017,syassen2008,weiner1999,bouganne2020}. The quench dynamics of the isolated Bose-Hubbard chain has already been extensively studied theoretically and numerically using for instance tensor network or variational Monte-Carlo techniques ~\cite{calabrese2006,barmettler2012,kollath2007,moeckel2008,manmana2009,roux2010,navez2010,carleo2014,krutitsky2014,despres2018,despres2019,sanchezpalencia2019,villa2020}. However, the quench dynamics associated to the latter 
quantum model in the presence of dissipations, corresponding here to local two-body losses, has been under-researched analytically and numerically~\cite{despresnew,liu2022,liu2024,nagao2024}. Most importantly, the possibility to perform quench spectroscopy to dissipative quantum lattice models has not been studied. The latter statement is also valid for non-Hermitian quantum lattice models. This exotic class of quantum systems appear naturally when studying the dynamics of dissipative quantum lattice models in the so-called no-click limit~\cite{daley2014}. Consequently, we also propose here to investigate the quench spectroscopy of the non-Hermitian ($s = 1/2$ short-range interacting) transverse-field Ising chain confined in the paramagnetic phase. Finally, we extend the quench spectroscopy method to isolated Hermitian quantum lattice models. More precisely, we show that the quench spectroscopy method is suitable and reliable for any quadratic quantum lattice model being bosonic or fermionic. To do so, the case study of the transverse-field Ising chain confined in the $z$-polarized phase is considered. Its quench dynamics induced by a sudden modification of the external field as well as its quench spectral function are investigated analytically using the bosonic and fermionic reformulations of its Hamiltonian in the latter gapped quantum phase while comparing the equation-of-motion and quasiparticle theoretical approaches. The previous statements underline the importance of the research work presented here together with the possibility to confirm our conclusions experimentally; at least for dissipative and isolated Hermitian quantum lattice models. \\

This article is organized as follows: in Sec.~\ref{sec:obh}, we start by introducing the dissipative one-dimensional Bose-Hubbard model and the quench procedure in Subsec.~\ref{subsec:model}. In Subsec.~\ref{subsec:benchmark}, we benchmark the equation-of-motion and quasiparticle approaches to unravel the main dynamical features of the isolated Bose-Hubbard chain. In Subsec.~\ref{subsec:double}, we discuss the dissipative and interaction quench dynamics of the Bose-Hubbard chain where the loss processes correspond here to on-site (local) two-body losses. In Subsec.~\ref{subsec:spectro}, we investigate theoretically the quench spectroscopy of the latter dissipative quantum lattice model. In Sec.~\ref{feasibility}, we discuss the experimental feasibility as well as the limitations associated to the quench spectroscopy method. Then, in Sec.~\ref{applicability}, the quench spectroscopy approach is extended to non-Hermitian quantum lattice models by considering the non-Hermitian transverse-field Ising chain confined in the paramagnetic phase. In Sec.~\ref{sec:srti}, we move on to the Hermitian version of the latter spin model to validate the quench spectroscopy method to any quadratic quantum lattice model being bosonic or fermionic. This model has the advantage to be diagonalized using a bosonic or fermionic reformulation discussed in Subsec.~\ref{subsec:boson} and Subsec.~\ref{subsec:fermion} respectively. Finally in Sec.~\ref{sec:conclusions}, we present our conclusions and possible extensions of the presented research work. 

\section{Dissipative quench dynamics of the Bose-Hubbard chain induced by local two-body losses}
\label{sec:obh}
\subsection{Model and quench procedure}
\label{subsec:model}

We focus on the one-dimensional Bose-Hubbard (1D BH) model for a lattice chain of length $L$ whose lattice spacing is fixed to unity, i.e. $a = 1$,
with periodic boundary conditions; for simplicity, we also set $\hbar = 1$. The corresponding Hamiltonian $\hat{H}$ reads:
\begin{equation}
\label{bhm}
\hat{H} = -J \sum_{R} \left(\hat{b}^{\dag}_{R} \hat{b}_{R+1} + \mathrm{h.c.}\right)+\frac{U}{2}\sum_R\hat{n}_{R}(\hat{n}_{R}-1),
\end{equation}

\noindent
where $\hat{b}_R$ and $\hat{b}_{R}^{\dag}$ correspond to the bosonic annihilation and creation operators acting on the lattice site $R \in \mathbb{N}$,
$\hat{n}_R = \hat{b}^{\dag}_R \hat{b}_R$ denotes the local occupation number operator associated to the lattice site index $R$, $J > 0$ corresponds to the hopping amplitude, $U>0$ is the on-site repulsive two-body interaction strength. At equilibrium and zero-temperature, the quantum phase diagram of the BH chain has been extensively studied~\cite{sachdev2001,cazalilla2011}. The latter displays a gapless superfluid (SF) phase and a gapped Mott-insulating (MI) phase with the so-called Mott lobes, determined by the competition between the hopping, the interactions and the chemical potential $\mu$ or equivalently the average filling $\bar{n}$ depending if the grand canonical or canonical ensemble is considered. In what follows, we rely on the average filling where for $\bar{n} \in \mathbb{N}^*$, the SF-MI phase transition is of the Berezinskii-Kosterlitz-Thouless type at the critical value $(U/J)_{\mathrm{c}} \simeq 3.3$ for $\bar{n} = 1$~\cite{kuhner2000one,kashurnikov1996exact,ejima2011,rombouts2006}. For non-integer fillings, the quantum system remains in the SF phase for any value of the dimensionless interaction parameter $U/J$. \\

The dissipative quench dynamics associated to the BH chain is fully characterized by the Lindblad master equation which reads as:
\begin{align}
\frac{\mathrm{d}}{\mathrm{d}t}\hat{\rho}= -i \left[\hat{H},\hat{\rho}\right] + \sum_{R} \hat{L}_{R} \hat{\rho} \hat{L}^{\dag}_{R} - \frac{1}{2} \left \{ \hat{L}^{\dag}_{R} \hat{L}_{R}, \hat{\rho}\right \},
\label{lind_mast_eq}
\end{align}
\noindent
where $\hat{H}$ refers to the Hamiltonian of the BH chain defined at Eq.~\eqref{bhm}. $\hat{\rho}$ refers to the time-dependent density matrix and $\hat{L}_{R}$ to the Lindblad jump operator acting on the lattice site $R$. For on-site two-body losses, $\hat{L}_R = \sqrt{\gamma} \hat{b}_R^2$ where $\gamma$ corresponds to the dissipation strength also called dissipation rate.\\

To drive the BH chain out of equilibrium, the following quench procedure is considered. We start from an initial many-body quantum state corresponding to the ground state of the 1D BH model without dissipation implying $\gamma = 0$. Initially, the density or equivalently the filling since $a = 1$ is unitary, i.e. $\bar{n} = 1$. Then, we let the quantum lattice model evolve with a non-zero value of the dissipation strength $\gamma > 0$. The latter defines the first quench corresponding to a dissipation quench. The second one is performed on the dimensionless interaction parameter $U/J$ hence corresponding to an interaction quench. It is done by considering a pre-quench value $U_{\mathrm{i}}/J$ quenched to a post-quench value $U_{\mathrm{f}}/J$. Note that the hopping amplitude $J$ is fixed during the full real time evolution. 
The double quench dynamics is studied starting from a many-body ground state of the BH chain confined in the SF phase, i.e. $U/J < (U/J)_{\mathrm{c}}$. To characterize the system's dynamics, we investigate a variety of quantum observables including $G_1(R,t) = \langle \hat{b}_R^{\dag}(t) \hat{b}_0(t) \rangle_c$ the connected one-body correlation function and $G_2(R,t) = \langle \hat{n}_R(t) \hat{n}_0(t) \rangle_c$ the connected density-density correlation function. The phase and density fluctuations determined by $G_1$ and $G_2$ can be measured in ultra-cold-atom experiments using time-of-flight and fluorescence microscopy imaging respectively ~\cite{cheneau2012,trotzky2012,langen2013,geiger2014}. \\

To investigate theoretically the dissipative quench dynamics induced by on-site two-body losses of the BH chain initially confined in the SF-mean-field regime, we rely on an analytical result from Ref.~\cite{despresnew}. In the latter paper, the set of equations of motion (EoMs) associated to the correlators $G_{\mathbf{k}}(t) = \langle \hat{b}^{\dag}_{\mathbf{k}} \hat{b}_{\mathbf{k}} \rangle_t = \langle \hat{n}_{\mathbf{k}} \rangle_t$ and $F_{\mathbf{k}}(t) = \langle \hat{b}_{-\mathbf{k}} \hat{b}_{\mathbf{k}} \rangle_t$ for the BH model on a $D$-dimensional hypercubic lattice for long-range two-body losses has been calculated. Using the Lindblad master equation, the EoMs are given by:
\begin{widetext}
\begin{subequations}
\label{lindblad_nl}
\begin{align}
& \frac{\mathrm{d}}{\mathrm{d}t} G_{\mathbf{k}}(t) = i \left \langle \left[\hat{H}(t), \hat{n}_{\mathbf{k}} \right] \right \rangle_t + \frac{1}{2}\sum_{\mathbf{R},\mathbf{R}'} \left( \left \langle \hat{L}_{\mathbf{R},\mathbf{R}'}^{\dag}\left[\hat{n}_{\mathbf{k}}, \hat{L}_{\mathbf{R},\mathbf{R}'} \right] \right \rangle_t + \mathrm{h.c} \right); \\
& \frac{\mathrm{d}}{\mathrm{d}t} F_{\mathbf{k}}(t) = i \left \langle \left[\hat{H}(t), \hat{b}_{-\mathbf{k}}\hat{b}_{\mathbf{k}} \right] \right \rangle_t + \sum_{\mathbf{R},\mathbf{R}'} \langle \hat{L}_{\mathbf{R},\mathbf{R}'}^{\dag}\hat{b}_{-\mathbf{k}}\hat{b}_{\mathbf{k}} \hat{L}_{\mathbf{R},\mathbf{R}'} \rangle_t -\frac{1}{2} \langle \hat{b}_{-\mathbf{k}}\hat{b}_{\mathbf{k}} \hat{L}_{\mathbf{R},\mathbf{R}'}^{\dag}\hat{L}_{\mathbf{R},\mathbf{R}'} \rangle_t -\frac{1}{2} \langle \hat{L}_{\mathbf{R},\mathbf{R}'}^{\dag}\hat{L}_{\mathbf{R},\mathbf{R}'} \hat{b}_{-\mathbf{k}}\hat{b}_{\mathbf{k}} \rangle_t; \\
& \hat{H}(t) = \frac{1}{2} \sum_{\mathbf{k} \neq \mathbf{0}}\mathcal{A}_{\mathbf{k}}(t) \left(\hat{b}^{\dag}_{\mathbf{k}} \hat{b}_{\mathbf{k}}+\hat{b}_{-\mathbf{k}}\hat{b}^{\dag}_{-\mathbf{k}}\right)+\mathcal{B}_{\mathbf{k}}(t) \left(\hat{b}^{\dag}_{\mathbf{k}} \hat{b}^{\dag}_{-\mathbf{k}}+\hat{b}_{\mathbf{k}} \hat{b}_{-\mathbf{k}} \right),
\label{bose_form_h_t}
\end{align}
\end{subequations}
\end{widetext}
\noindent
where $\hat{H}(t)$ denotes the quadratic form in the reciprocal Fourier space of the BH model in the SF-mean-field regime with a time-dependent filling $\bar{n}(t)$ or equivalently a time-dependent density $n(t)$. The latter is found using a standard mean-field approximation~\cite{roux2013,despres2019}. The momentum- and time-dependent functions $\mathcal{A}_{\mathbf{k}}(t)$ and $\mathcal{B}_{\mathbf{k}}(t)$ depending on $n(t)$, or equivalently on $\bar{n}(t)$, are defined later on. $\hat{b}_{\mathbf{k}}$ and $\hat{b}_{\mathbf{k}}^{\dag}$ correspond to the bosonic annihilation and creation operators acting in the reciprocal space on the momentum $\mathbf{k}$; $\hat{n}_{\mathbf{k}} = \hat{b}^{\dag}_{\mathbf{k}} \hat{b}_{\mathbf{k}}$ denotes the occupation number operator in Fourier space for the momentum $\mathbf{k}$. The previous set of equations is characterized by the Lindblad jump operator $\hat{L}_{\mathbf{R},\mathbf{R}'}$ as well as the long-range dissipation strength $\gamma_{||\mathbf{R}-\mathbf{R'}||}$ which read as:
\begin{equation}
\hat{L}_{\mathbf{R},\mathbf{R}'} = \sqrt{\gamma_{||\mathbf{R}-\mathbf{R}'||}} \hat{b}_{\mathbf{R}} \hat{b}_{\mathbf{R}'},~\gamma_{||\mathbf{R}-\mathbf{R}'||} = \frac{\Gamma}{(1+||\mathbf{R}-\mathbf{R}'||)^{\alpha}},
\end{equation}
\noindent
with $||.||$ standing for the vector norm and $\alpha$ for the power-law exponent characterizing the spatial decay of the long-range two-body loss processes. The Hermiticity of the occupation number operator in momentum space $\hat{n}_{\mathbf{k}}$ has been considered to simplify the EoM associated to the correlator $G_{\mathbf{k}}(t)$. To calculate the EoMs while considering the SF-mean-field regime, a decoupling of the condensate mode $\mathbf{k} = \mathbf{0}$ from the other modes is used together with a mean-field approximation where only the terms depending on correlators involving four or two bosonic operators acting on the mode $\mathbf{k} = \mathbf{0}$ are conserved. As a consequence, the EoMs associated to $G_{\mathbf{k}}(t)$ and $F_{\mathbf{k}}(t)$ both for $\mathbf{k} = \mathbf{0}$ and $\mathbf{k} \neq \mathbf{0}$ have been calculated. Finally, the following set of EoMs is found:
\begin{widetext}
\begin{subequations}
\label{EoMs_nl}
\begin{align}
\frac{\mathrm{d}}{\mathrm{d}t} G_{\mathbf{0}}(t) =&~ -\sum_{\mathbf{q} \neq \mathbf{0}} \left[\left(\mathcal{G}_{\mathbf{q}}+\mathcal{H}_{\mathbf{q}}\right)F_{\mathbf{0}}(t)F_{\mathbf{q}}(t)^* + \mathrm{h.c} \right] -\sum_{\mathbf{q} \neq \mathbf{0}} \mathcal{F}_{\mathbf{q}}G_{\mathbf{0}}(t)G_{\mathbf{q}}(t) - 2\mathcal{G}_{\mathbf{0}}G_{\mathbf{0}}(t)(G_{\mathbf{0}}(t)-1); \\
\frac{\mathrm{d}}{\mathrm{d}t} F_{\mathbf{0}}(t) =&~ - \mathcal{G}_{\mathbf{0}}(2G_{\mathbf{0}}(t)-3)F_{\mathbf{0}}(t) -\sum_{\mathbf{q} \neq \mathbf{0}}\mathcal{F}_{\mathbf{q}}F_{\mathbf{0}}(t)G_{\mathbf{q}}(t) - \sum_{\mathbf{q} \neq \mathbf{0}}\left(\mathcal{G}_{\mathbf{q}}+\mathcal{H}_{\mathbf{q}}\right)(2G_{\mathbf{0}}(t)+1)F_{\mathbf{q}}(t); \\
\frac{\mathrm{d}}{\mathrm{d}t} G_{\mathbf{k}}(t) =&~ -2\mathcal{B}_{\mathbf{k}}(t)\operatorname{Im}(F_{\mathbf{k}}(t)) - \mathcal{G}_{\mathbf{k}}(F_{\mathbf{0}}(t)F_{\mathbf{k}}(t)^* + \mathrm{h.c}) - \mathcal{F}_{\mathbf{k}}G_{\mathbf{0}}(t)G_{\mathbf{k}}(t)~~~~\forall \mathbf{k} \neq \mathbf{0}; \\
\frac{\mathrm{d}}{\mathrm{d}t} F_{\mathbf{k}}(t) =&~ -\left[2i\mathcal{A}_{\mathbf{k}}(t) + \mathcal{F}_{\mathbf{k}}G_{\mathbf{0}}(t)\right]F_{\mathbf{k}}(t) -\left[i\mathcal{B}_{\mathbf{k}}(t) + \mathcal{G}_{\mathbf{k}} F_{\mathbf{0}}(t) \right](2G_{\mathbf{k}}(t)+1)~~~~\forall \mathbf{k} \neq \mathbf{0},
\end{align}
\end{subequations}
\end{widetext}
\noindent
where the momentum- and possibly time-dependent functions $\mathcal{A}_{\mathbf{k}}(t)$, $\mathcal{B}_{\mathbf{k}}(t)$, $\mathcal{F}_{\mathbf{q}}$, $\mathcal{G}_{\mathbf{q}}$ and $\mathcal{H}_{\mathbf{q}}$ are defined as follows:
\begin{widetext}
\begin{subequations}
\begin{align}
&\mathcal{A}_{\mathbf{k}}(t) = 4J\sum_{i=1}^{D} \sin^2\left(\frac{\mathbf{k} \cdot \mathbf{d}_i}{2}\right) + \mathcal{B}_{\mathbf{k}}(t),~~ \mathcal{B}_{\mathbf{k}}(t) = ~ \frac{U}{L^D} \sum_{\mathbf{q}} G_{\mathbf{q}}(t) = U\bar{n}(t);\\
&\mathcal{F}_{\mathbf{q}} = 2(\mathcal{G}_{\mathbf{0}}+\mathcal{G}_{\mathbf{q}}), ~~\mathcal{G}_{\mathbf{q}} = \frac{1}{L^{2D}}\sum_{\mathbf{R},\mathbf{R}'}\gamma_{||\mathbf{R}-\mathbf{R}'||}\cos[\mathbf{q}(\mathbf{R}-\mathbf{R}')],~~
\mathcal{H}_{\mathbf{q}} = \frac{i}{L^{2D}}\sum_{\mathbf{R},\mathbf{R}'}\gamma_{||\mathbf{R}-\mathbf{R}'||}\sin[\mathbf{q}(\mathbf{R}-\mathbf{R}')],
\end{align}
\end{subequations}
\end{widetext}
\noindent
with $\mathbf{k} = (k_x, k_y, k_z, ...)^{T}$, $\mathbf{d}_i = (0, ..., 0, a_i, 0, ..., 0)^{T}$ with $a_i = 1$ denoting the lattice spacing along the $i$-th spatial dimension and $i \in [|1,D|]$. For our case study corresponding to the BH model on a 1D lattice chain, i.e. $D = 1$, while considering on-site two-body losses, i.e. 
$\alpha \rightarrow + \infty$ leading to $\gamma_{R-R'} = \Gamma \delta_{R,R'} = \gamma \delta_{R,R'}$, the Lindblad jump operator becomes $\hat{L}_{R} = \sqrt{\gamma} \hat{b}_{R}^2$. It also yields for the momentum-dependent function:
\begin{widetext}
\begin{subequations}
\begin{align}
& \mathcal{A}_{\mathbf{k}}(t) = 4J\sin^2(k/2) + \mathcal{B}_{k}(t),~~~ \mathcal{B}_{\mathbf{k}}(t) = \frac{U}{L} \sum_q G_q(t) = U \bar{n}(t),~~~ \mathcal{F}_{\mathbf{q}} = \frac{4\gamma}{L},~~~\mathcal{G}_{\mathbf{q}} = \frac{\gamma}{L},~~~ \mathcal{H}_{\mathbf{q}} = 0.
\end{align}
\end{subequations}
\end{widetext}
\noindent
Finally, the set of EoMs at Eq.~\eqref{EoMs_nl} can be simplified using the previous relations and it yields: 
\begin{widetext}
\begin{subequations}
\label{EoMs}
\begin{align}
\frac{\mathrm{d}}{\mathrm{d}t} G_0(t) =&~ -\frac{\gamma}{L}\sum_{q \neq 0}(F_0(t)F_q(t)^* + \mathrm{h.c}) - \frac{4\gamma}{L}G_0(t)\sum_{q \neq 0}G_q(t) - \frac{2\gamma}{L}G_0(t)\left(G_0(t)-1\right); \\
\frac{\mathrm{d}}{\mathrm{d}t} F_0(t) =&~  -\frac{\gamma}{L}\left(2G_0(t)-3\right)F_0(t) - \frac{4\gamma}{L}F_0(t)\sum_{q\neq 0}G_q(t) - \frac{\gamma}{L}(2G_0(t)+1)\sum_{q\neq 0} F_q(t); \\
\frac{\mathrm{d}}{\mathrm{d}t} G_k(t) =&~  -2\mathcal{B}_{k}(t)\operatorname{Im}(F_k(t)) - \frac{\gamma}{L}\left( F_0(t)F_k(t)^* + \mathrm{h.c} \right) -\frac{4\gamma}{L}G_0(t)G_k(t),~ \forall k\neq 0; \\
\frac{\mathrm{d}}{\mathrm{d}t} F_k(t) =&~ -\left[2i\mathcal{A}_{k}(t)+\frac{4\gamma}{L}G_0(t)\right]F_k(t) - \left[i\mathcal{B}_{k}(t)+ \frac{\gamma}{L}F_0(t)\right](2G_k(t)+1),~ \forall k \neq 0,
\end{align}
\end{subequations}
\end{widetext}
\noindent
where the initial conditions for the condensate mode $k = 0$ and the modes $k \neq 0$ are given by: 
\begin{subequations}
\label{eq_init}
\begin{align}
G_0(0) =&~ N_0 = N - \sum_{k \neq 0} G_k(0); \\
F_0(0) =&~ \Theta(U) N_0; \\
G_k(0) =&~ \frac{1}{2} \left(\frac{\mathcal{A}_{k}}{\mathcal{E}_{k}} - 1 \right); \\
F_k(0) =&~ -\frac{\mathcal{B}_{k}}{2\mathcal{E}_{k}},
\end{align}
\end{subequations}
\noindent
with $\Theta(U)$ denoting the Heaviside function defined as $\Theta(U) = 1$ if $U > 0$ and $\Theta(U) = 0$ if $U = 0$. The low-lying excitation spectrum $\mathcal{E}_{k}$ and the coefficients $\mathcal{A}_{k}$ and $\mathcal{B}_{k}$ are given by:
\begin{align}
\label{eq_spectrum}
&\mathcal{E}_{k} = \sqrt{\mathcal{A}_{k}^2 - \mathcal{B}_{k}^2},~~~ \mathcal{A}_{k} = 4J\sin^2(k/2) + \mathcal{B}_{k},~~~\mathcal{B}_{k} = U\bar{n}.
\end{align}

\subsection{Quench dynamics of the isolated Bose-Hubbard chain}
\label{subsec:benchmark}
In Ref.~\cite{despresnew}, the validity of the EoM approach to describe accurately the quench dynamics of the dissipative BH chain characterized by on-site (or long-range) two-body losses and initially confined in the SF-mean-field regime has been certified against tensor networks numerical simulations. Hence, we can rely on the latter to describe the quench dynamics of the isolated BH chain in this same gapless phase for a sudden global quench on the interaction strength $U$. However, due to this new framework, the theory requires several adjustments. In what follows, we present them and benchmark the theoretical predictions of the EoM approach with those provided by a quasiparticle theory proposed in Ref.~\cite{despres2018, despres_these}. \\

In what follows, a sudden global quench on the interaction strength $U$ is performed for the isolated BH chain confined in the SF-mean-field regime. To do so, the set of EoMs has to be slightly modified: \\

$\tiny \bullet~$ The quantum system being isolated has $U(1)$ symmetry, i.e. particle-number conservation, implying that the momentum- and time-dependent functions $\mathcal{A}_{k}(t)$ and $\mathcal{B}_{k}(t)$ depending on $n(t)$ remain constant in time and become equal to their initial value.\\

$\tiny\bullet~$ The coefficients $\mathcal{A}_{k}$ and $\mathcal{B}_{k}$ in the EoMs have to be replaced by $\mathcal{A}_{k,\mathrm{f}}$ and $\mathcal{B}_{k,\mathrm{f}}$ taking into account the post-quench value of $U$ 
denoted by $U_{\mathrm{f}}$.\\

$\tiny \bullet~$ Similarly for the initial conditions, $\mathcal{A}_{k}$ and $\mathcal{B}_{k}$ are replaced by $\mathcal{A}_{k,\mathrm{i}}$ and $\mathcal{B}_{k,\mathrm{i}}$ depending on the pre-quench value $U_{\mathrm{i}}$.\\

$\tiny \bullet~$  $\gamma = 0$ for each EoM in order to remove dissipation effects. \\

\begin{figure}
\centering
\includegraphics[scale = 0.41]{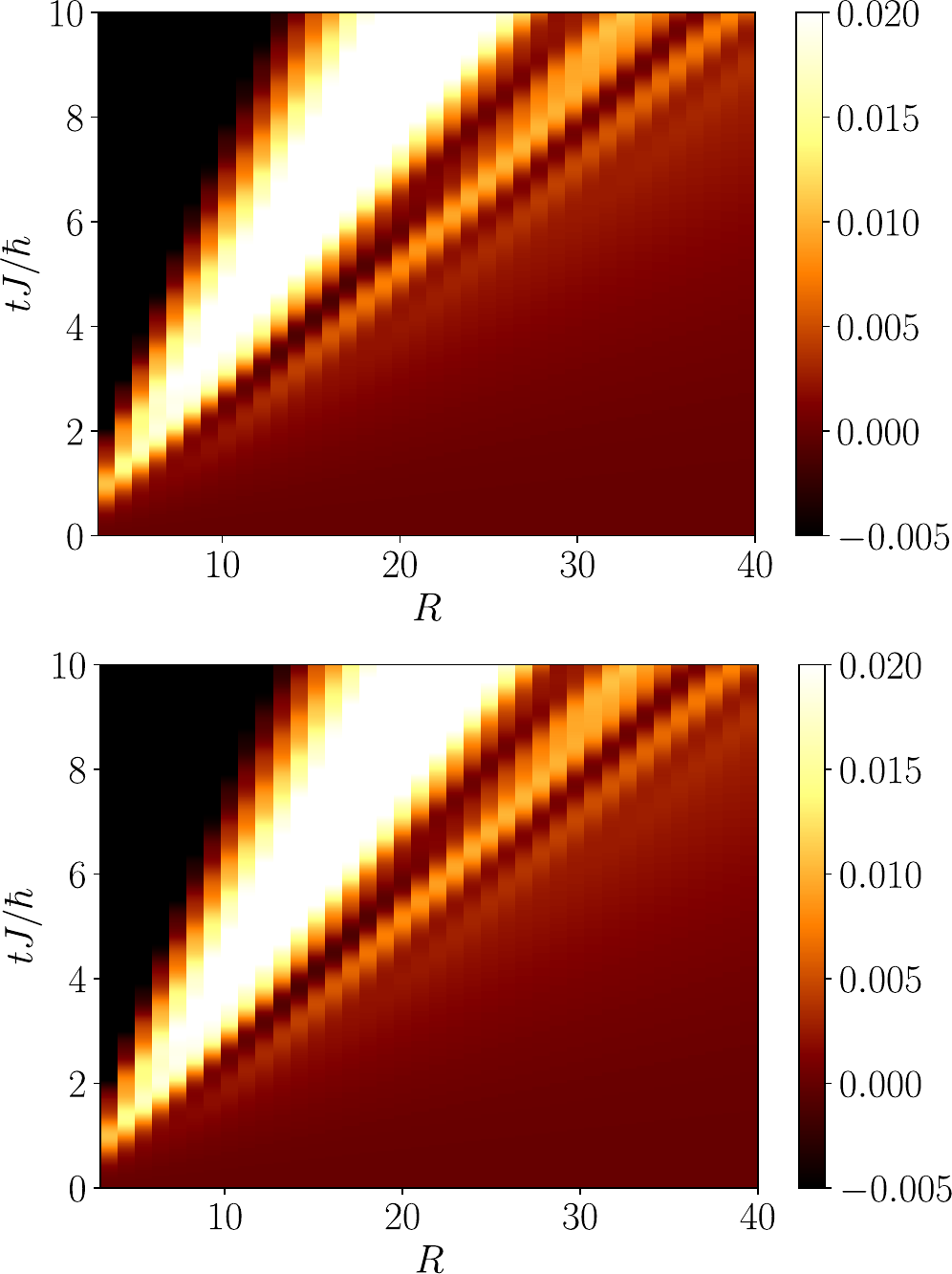}
\caption{Connected one-body correlation function $G_1(R,t)$ for a sudden global quench on $U$ of the BH chain confined in the SF-mean-field regime. On the top panel, the QP approach is considered and on the bottom panel, the EoM approach is used. The considered dimensionless interaction parameters are: $U_{\mathrm{f}}\bar{n}/J = 0.5$, $U_{\mathrm{i}}\bar{n}/J = 1$.}
\label{fig_g1_closed}
\end{figure}

\noindent
Finally, we get the following set of EoMs:
\begin{subequations}
\label{EoM_isolated}
\begin{align}
\frac{\mathrm{d}}{\mathrm{d}t} G_k(t) =&~ -2\mathcal{B}_{k,\mathrm{f}}\operatorname{Im}(F_k(t)); \\
\frac{\mathrm{d}}{\mathrm{d}t} F_k(t) =&~ -2i\mathcal{A}_{k,\mathrm{f}}F_k(t) - i\mathcal{B}_{k,\mathrm{f}}(1+2G_k(t)),
\end{align}
\end{subequations}

\noindent
where the initial conditions are defined as follows:
\begin{align}
G_k(0) =&~ \frac{1}{2} \left(\frac{\mathcal{A}_{k,\mathrm{i}}}{\mathcal{E}_{k,\mathrm{i}}} - 1 \right),~~F_k(0) = -\frac{\mathcal{B}_{k,\mathrm{i}}}{2\mathcal{E}_{k,\mathrm{i}}},
\end{align}

\noindent
with the momentum-dependent pre-quench functions $\mathcal{A}_{k,\mathrm{i}}$ and $\mathcal{B}_{k,\mathrm{i}}$ and the pre-quench quasiparticle dispersion relation 
$\mathcal{E}_{k,\mathrm{i}}$ having the following form:
\begin{subequations}
\begin{align}
\mathcal{E}_{k,\mathrm{i}} =&~ \sqrt{\mathcal{A}_{k,\mathrm{i}}^2 - \mathcal{B}_{k,\mathrm{i}}^2},\\
\mathcal{A}_{k,\mathrm{i}} =&~4J\sin^2(k/2) + U_{\mathrm{i}}\bar{n}, \\
\mathcal{B}_{k,\mathrm{i}} =&~ U_{\mathrm{i}}\bar{n}.
\end{align}
\end{subequations}

\noindent
Note that if no quench is induced, i.e. $U_{\mathrm{i}} = U_{\mathrm{f}}$, then the EoMs are strictly equal to zero and we recover $F_k(t) = F_k(0)$ and $G_k(t) = G_k(0)$ as expected. \\

In what follows, the connected one-body correlation function $G_1(R,t)$ defined as $G_1(R,t) = \langle \hat{b}^{\dag}_R \hat{b}_0 \rangle_t - \langle \hat{b}^{\dag}_R \hat{b}_0 \rangle_0$ is considered. In reciprocal space, the latter reads:
\begin{align}
& G_1(R,t) = \frac{1}{L} \sum_{k} e^{ikR}\left( \langle \hat{n}_k \rangle_t - \langle \hat{n}_k \rangle_0 \right).
\label{G1_th}
\end{align}
\noindent
The latter can be further simplified. The expression of $\hat{H}_{\mathrm{f}}$ in reciprocal space is given at Eq.~\eqref{bose_form_h_t} where $\mathcal{A}_k(t)$ and $\mathcal{B}_k(t)$ are respectively replaced by $\mathcal{A}_{k,\mathrm{f}}$ and 
$\mathcal{B}_{k,\mathrm{f}}$. The time dependence vanishes due to the particle-number conservation since the isolated BH chain is considered, see also Refs.~\cite{despres2018,despres_these,despresnew}. From the latter, we notice that the condensate mode $k = 0$ has been integrated out using the standard mean-field approximation and thus leading to the filling (or equivalently the density) dependence $\bar{n}$ within the post-quench momentum-dependent functions $\mathcal{A}_{k,\mathrm{f}}$ and $\mathcal{B}_{k,\mathrm{f}}$. 
It immediately follows the property:
\begin{equation}
\frac{\mathrm{d}\langle \hat{n}_0 \rangle_t}{\mathrm{d}t} = i\langle [\hat{H}_{\mathrm{f}},\hat{n}_0] \rangle_t = 0,
\end{equation}

\noindent
meaning that $\langle \hat{n}_0 \rangle_t = \langle \hat{n}_0 \rangle_0$. Then, $G_1(R,t)$ reduces to:
\begin{align}
& G_1(R,t) = \frac{1}{L}\sum_{k\neq0} \cos(kR) \left[G_k(t)-G_k(0)\right],
\label{G1_isolated}
\end{align}
\noindent
where $G_k(t) = \langle \hat{n}_k \rangle_t = \langle \hat{b}^{\dag}_k \hat{b}_k \rangle_t$. Our EoM approach is compared to a quasiparticle (QP) approach based on the Heisenberg picture together with the introduction of Bogolyubov quasiparticle operators~\cite{despres2018,despres_these}. Its validity has been certified numerically using tensor network based techniques ~\cite{despres2019}. The QP approach permits to find an explicit expression of $G_1(R,t)$ which reads as: 
\begin{align}
G_1(R,t) =&~ \frac{1}{L} \sum_{k \neq 0} \mathcal{S}^{(1)}_k \cos(kR) \sin^2(\mathcal{E}_{k,\mathrm{f}}t),
\end{align}
\noindent
where the amplitude function is denoted by $\mathcal{S}^{(1)}_k$ and defined as:
\begin{align}
\mathcal{S}^{(1)}_k =&~ \frac{\left(\mathcal{A}_{k,\mathrm{i}}\mathcal{B}_{k,\mathrm{f}} - \mathcal{A}_{k,\mathrm{f}}\mathcal{B}_{k,\mathrm{i}}\right)\mathcal{B}_{k,\mathrm{f}}}{\mathcal{E}_{k,\mathrm{i}}\mathcal{E}_{k,\mathrm{f}}^2}.
\end{align}

\noindent
On Fig.~\ref{fig_g1_closed}, both theoretical approaches are compared and we find a perfect agreement between them. This validates the EoM approach defined by the set of coupled differential equations at Eq.~\eqref{EoMs}. Note that both theories are equivalent since they are based on the same expression of the diagonalized form of the Hamiltonian $\hat{H}$ in Fourier space. While the EoM approach is based on the calculation of coupled first-order differential equations of quadratic correlators, the QP approach directly calculates the expression of the time-evolved Bogolyubov quasiparticle operators and thus permits to find an explicit analytical expression of the correlation functions. \\

Regarding the density fluctuations characterized via the $G_2$ connected density-density correlation function, the latter may be written as follows:
\begin{figure}
\centering
\includegraphics[scale = 0.41]{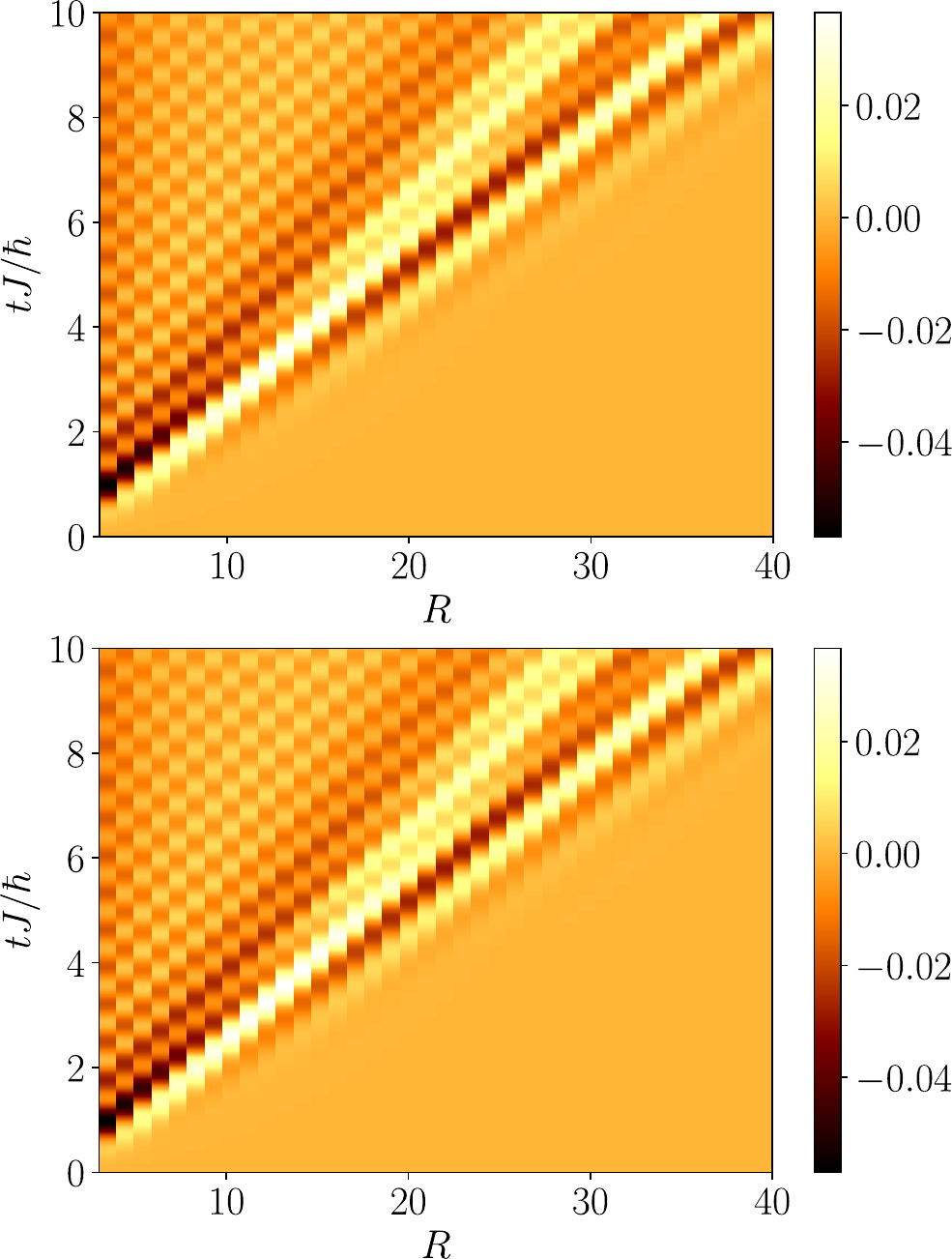}
\caption{Connected density-density correlation function $G_2(R,t)$ for a sudden global quench on $U$ of the BH chain confined in the SF-mean-field regime. On the top panel, the QP approach is considered and on the bottom panel, the EoM approach is used. The considered dimensionless interaction parameters are: $U_{\mathrm{f}}\bar{n}/J = 0.5$, $U_{\mathrm{i}}\bar{n}/J = 1$.}
\label{fig_g2_closed}
\end{figure}

\begin{align}
& G_{2}(R,t) = \langle \hat{n}_{R} \hat{n}_{0} \rangle_t - \langle \hat{n}_{R} \rangle_t \langle \hat{n}_{0} \rangle_t - \langle \hat{n}_{R} \hat{n}_{0} \rangle_0 + \langle \hat{n}_{R} \rangle_0 \langle \hat{n}_{0} \rangle_0.
\end{align}

\noindent
Relying on the mean-field approximation, $G_2$ reads as:
\begin{align}
& G_2(R,t) = \frac{2\bar{n}}{L} \sum_{k\neq 0} \cos(kR) \{ \operatorname{Re}[F_k(t)] + G_k(t) \nonumber\\
& - \operatorname{Re}[F_k(0)] - G_k(0)\},
\label{eq_G_2}
\end{align}

\noindent
Note that $G_1(R,0) = G_2(R,0) = 0$ as expected since both $G_1$ and $G_2$ refer to the connected version of the one-body and density-density correlation function respectively. The previous analytical form of $G_2(R,t)$ is consistent with 
the following expression found using the set of EoMs defined at Eq.~\eqref{EoMs}:
\begin{align}
\label{theory_g2}
& g_2(R,t) = n(t)^2 + \frac{2n(t)}{L} \sum_{k\neq 0} \cos(kR) \left\{\operatorname{Re}[F_k(t)] + G_k(t)\right\},
\end{align}

\noindent
where $g_2(R,t) = \langle \hat{n}_{R} \hat{n}_{0} \rangle_t$ denotes the non-connected density-density correlation function in the SF-mean-field regime while considering loss processes via a time-dependent density $n(t)$. The latter expression is discussed in Appendix~\ref{g2} together with the theoretical expression of the condensate density $n(t)$ in the non-interacting case $U = 0$. By considering particle-number conservation, i.e. $n(t) = n = \bar{n}$, for the isolated BH chain and according to the definition of $G_2$ and $g_2$, we can deduce that $G_2(R,t) = g_2(R,t) - g2(R,0)$. Hence, we find:
\begin{align}
& G_2(R,t) = \bar{n}^2 + \frac{2\bar{n}}{L} \sum_{k\neq 0} \cos(kR) \left[\operatorname{Re}(F_k(t)) + G_k(t)\right] \nonumber\\
&- \bar{n}^2 - \frac{2\bar{n}}{L} \sum_{k\neq 0} \cos(kR) \left[\operatorname{Re}(F_k(0)) + G_k(0)\right],
\end{align}

\noindent
leading finally to the expected expression of $G_2$ at Eq.~\eqref{eq_G_2}. Note that the connected part of the $G_2$ correlation function namely 
$-\langle \hat{n}_{R} \rangle_t \langle \hat{n}_{0} \rangle_t + \langle \hat{n}_{R} \rangle_0 \langle \hat{n}_{0} \rangle_0$ vanishes for the isolated BH chain. Indeed, 
according to Ref.~\cite{despresnew}, in the presence of particle losses the $G_1$ (non-connected) one-body correlation function also called one-body density matrix is defined and takes the following form in the SF-mean-field regime:
\begin{align}
& G_1(R,t) = \langle \hat{b}^{\dag}_{R} \hat{b}_{0} \rangle_t = n(t) = G_1(0,t) = \langle \hat{n}_{0} \rangle_t,
\end{align}

\noindent
where the mean-field approximation has been considered to deduce the latter form being spatially independent. Using the translational invariance of the model, we end up
with $\langle \hat{n}_{0} \rangle_t = \langle \hat{n}_{R} \rangle_t$, $\forall R \in \mathbb{N}$. Finally, for the isolated BH chain, we have 
$\langle \hat{n}_{0} \rangle_t = \langle \hat{n}_{R} \rangle_t = \bar{n} = \langle \hat{n}_{0} \rangle_0 = \langle \hat{n}_{R} \rangle_0$ and thus we recover that the connected part of $G_2$ vanishes as expected. \\

$G_2$ the equal-time connected density-density correlation function at Eq.~\eqref{eq_G_2} is deduced theoretically by injecting the quadratic bosonic correlators $F$ and $G$ by their values obtained by solving their respective EoM. The second theoretical approach is given by the QP theory, see Ref.~\cite{despres_these}, leading to: 
\begin{subequations}
\begin{align}
G_2(R,t) =&~ \frac{1}{L} \sum_{k\neq 0} \mathcal{S}^{(2)}_k \cos(kR) \sin^{2}(\mathcal{E}_{k,\mathrm{f}}t), \\
\mathcal{S}^{(2)}_k =&~ \frac{2\bar{n}\left(\mathcal{A}_{k,\mathrm{f}}\mathcal{B}_{k,\mathrm{i}} - \mathcal{A}_{k,\mathrm{i}}\mathcal{B}_{k,\mathrm{f}}\right)}{\left(\mathcal{A}_{k,\mathrm{f}} + \mathcal{B}_{k,\mathrm{f}}\right)\mathcal{E}_{k,\mathrm{i}}}.
\end{align}
\end{subequations}

\noindent
On Fig.~\ref{fig_g2_closed}, both theoretical approaches are compared and we find a perfect agreement between them. This validates for another observable the EoM approach defined by the set of coupled differential equations at Eq.~\eqref{EoMs}. Note that the set of EoMs at Eq.~\eqref{EoM_isolated} is valid when considering the Hamiltonian $\hat{H}$ of the BH chain confined in the SF-mean-field regime which can be expressed in a generic quadratic Bose form in
reciprocal space: 
\begin{equation}
\hat{H} = \frac{1}{2} \sum_{k \neq 0}\mathcal{A}_{k}\left(\hat{b}^{\dag}_{k} \hat{b}_{k}+\hat{b}_{-k}\hat{b}^{\dag}_{-k}\right)+\mathcal{B}_{k} \left(\hat{b}^{\dag}_{k} \hat{b}^{\dag}_{-k}+\hat{b}_{k} \hat{b}_{-k} \right),
\label{eq_bhm_quadratic}
\end{equation}

\noindent
with $\mathcal{A}_{k} = 4J\sin^2(k/2) + U\bar{n}$ and $\mathcal{B}_{k} = U \bar{n}$, see Eq.~\eqref{bose_form_h_t} and Refs.~\cite{roux2013,despres2019}. The latter can be reformulated in a Bogolyubov-de-Gennes
form and reads as:
\begin{align}
& \hat{H} = \frac{1}{2} \sum_{k \neq 0}
\begin{pmatrix}
\hat{b}^{\dag}_k & \hat{b}_{-k} 
\end{pmatrix} \hat{H}_{\mathrm{BdG}}(k) 
\begin{pmatrix}
\hat{b}_k \\
\hat{b}^{\dag}_{-k}
\end{pmatrix}; \\
& \hat{H}_{\mathrm{BdG}}(k) =
\begin{pmatrix}
\mathcal{A}_{k}& \mathcal{B}_{k} \\
\mathcal{B}_{k} & \mathcal{A}_{k} 
\end{pmatrix},
\end{align}

\noindent
where $\hat{H}_{\mathrm{BdG}}$ refers to the bosonic Bogolyubov-de-Gennes Hamiltonian. Similar statements will 
apply for any quantum system whose Hamiltonian can be reformulated in a quadratic Bose form, i.e. can be 
cast into the previous form. The latter statements are also valid for fermionic quantum systems provided that their Hamiltonian displays a quadratic Fermi form, i.e. can be expressed via a fermionic Bogolyubov-de-Gennes Hamiltonian. In Section~\ref{srti}, the latter expectations are discussed by considering the $s = 1/2$ short-range interacting transverse-field Ising chain in the paramagnetic phase. \\

By considering the thermodynamic limit, i.e. $L \rightarrow +\infty$, and by performing some trivial algebra, for both connected equal-time correlation functions can be expressed in the following generic form ~\cite{despres2018}: 
\begin{align}
&G_{\mathrm{i}}(R,t) \sim \int_{\mathcal{B}} \mathrm{d}k \mathcal{S}_k^{(\mathrm{i})} \left[ e^{ i(kR+2\mathcal{E}_{k,\mathrm{f}}t)}+ e^{i(kR-2\mathcal{E}_{k,\mathrm{f}}t)} \right],\label{generic}
\end{align}
\noindent
where $\mathcal{B} = [-\pi, \pi]$ denotes the first Brillouin zone. Note that the latter form unveiled at Ref.~\cite{despres2018} will be essential for the discussion about the quench spectroscopy method (QS) applied to the dissipative Bose-Hubbard chain. More precisely, Eq.~\eqref{generic} consists in the generic form of equal-time connected correlation functions (ETCCF) for an isolated quantum lattice model driven out of equilibrium via a weak sudden global quench. In Ref.~\cite{despres2018}, the large distance and time behavior of the latter expression is characterized using the stationary phase approximation to unravel general statements regarding the correlation spreading in closed quantum systems. For short-range interacting quantum models, the space-time pattern of ETCCFs
is expected to display a twofold linear structure. This double structure is characterized by a ballistic, i.e. linear, correlation edge (CE) separating the causal region of the correlations from the non-causal one. The causal region refers to the region where the correlations are different from zero. In other words, beyond 
the CE, the space-time correlations are exponentially suppressed. The second structure consists in the ballistic spreading of a series of local maxima and minima in the vicinity of the CE. The CE velocity is given by twice the maximal group velocity corresponding to the quasiparticle pair with the highest velocity, i.e.  $V_{\mathrm{CE}} = 2V_{\mathrm{g}}(k^*) = 2\mathrm{max}(\mathrm{d}\mathcal{E}_{k,\mathrm{f}}/\mathrm{d}k)$ with $k^* = \mathrm{argmax}(\mathrm{d}\mathcal{E}_{k,\mathrm{f}}/\mathrm{d}k)$. The velocity associated to the linear spreading of the local extremum is given by twice the phase velocity evaluated at the quasi-momentum for which the group velocity is maximal, i.e. $V_{\mathrm{m}} = 2V_{\varphi}(k^*) = 2 \mathcal{E}_{k^*}/k^*$. The validity of the latter behavior has been certified both analytically~\cite{despres_these, despres2018} and numerically using whether time-dependent matrix product state calculations~\cite{despres2019} or time-dependent variational Monte-Carlo calculations~\cite{carleo2014}.   

\subsection{Double quench dynamics of the Bose-Hubbard chain}
\label{subsec:double}
In what follows, we investigate the behavior of $G_2(R,t)$ for a sudden global quench not only on the dissipation strength from $\gamma = 0$ to $\gamma > 0$ but also on the interaction strength $U$ from $U = U_{\mathrm{i}}$ to $U = U_{\mathrm{f}}$ such that the BH chain is initially confined in the SF phase. Experimentally, the quench on $U$ can be realized in ultra-cold-atom experiments by suddenly lowering or raising the depth of the optical lattice in order to decrease or to increase the dimensionless interaction parameter $U/J$ respectively~\cite{cheneau2012}. In this context, the EoMs at Eq.~\eqref{EoMs} as well as the corresponding initial conditions at Eq.~\eqref{eq_init} have to be slightly modified. In the EoMs, $\mathcal{A}_{k}$ and $\mathcal{B}_{k}$ are replaced by $\mathcal{A}_{k,\mathrm{f}}$ and $\mathcal{B}_{k,\mathrm{f}}$ depending now on the post-quench interaction strength $U_{\mathrm{f}}$. For the initial conditions,
$\mathcal{A}_{k}$ and $\mathcal{B}_{k}$ are replaced by $\mathcal{A}_{k,\mathrm{i}}$ and $\mathcal{B}_{k,\mathrm{i}}$ depending now on the pre-quench interaction strength $U_{\mathrm{i}}$. $G_2$ the ETCCF to investigate density fluctuations is computed by solving the previous set of EoMs whose values of the relevant correlators are injected in the theoretical expression of $g_2$ given by:
\begin{align}
& g_2(R,t) = n(t)^2 + \frac{2n(t)}{L}\sum_{k \neq 0} \cos(kR)\left\{G_k(t) + \operatorname{Re}[F_k(t)]\right\}.
\label{eq_g2}
\end{align}

\begin{figure}[h!]
\centering
\includegraphics[scale = 0.38]{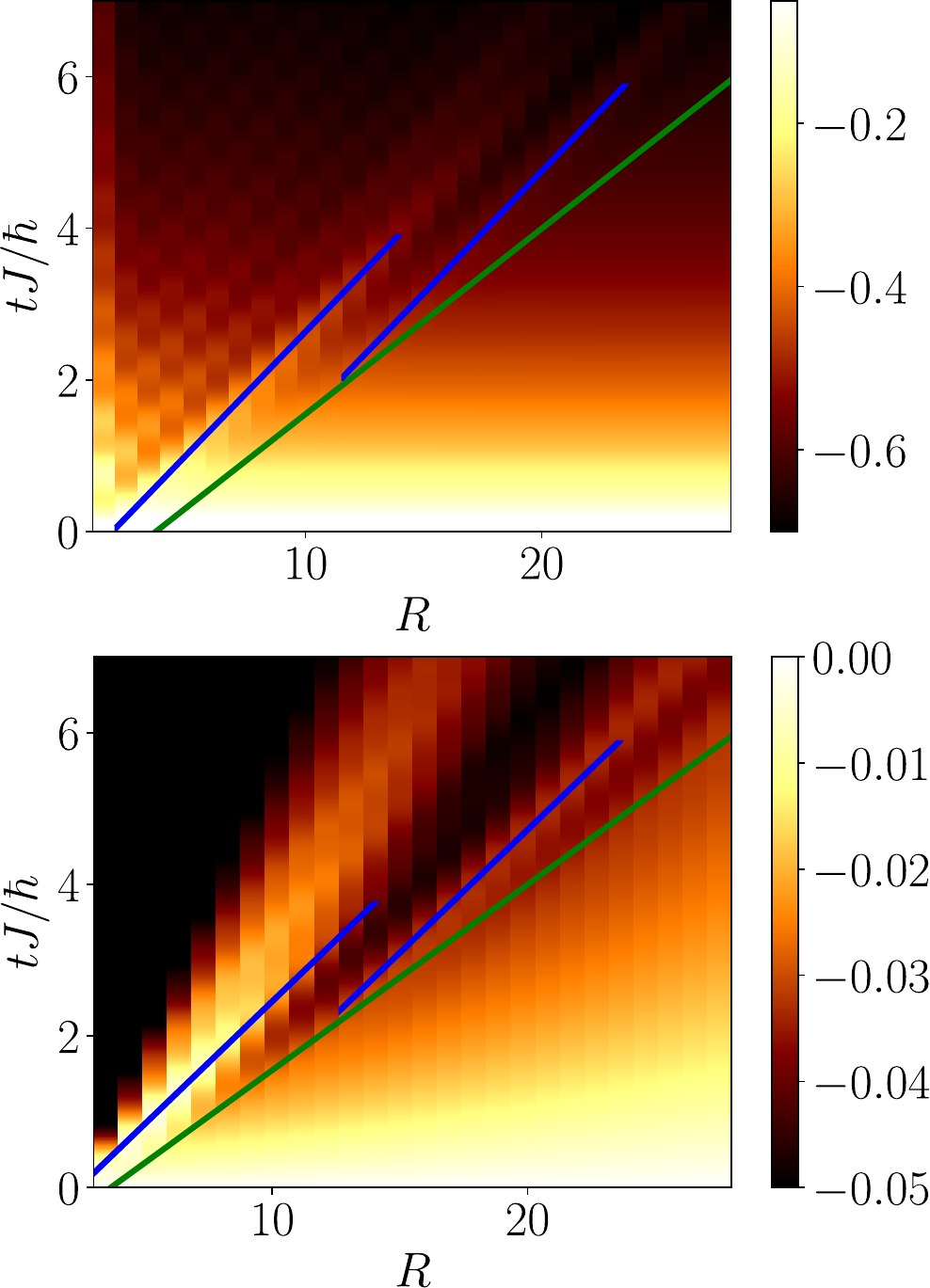}
\caption{Space-time pattern of (top panel) $G_2(R,t) = \langle \hat{n}_{R}(t) \hat{n}_{0}(t) \rangle_{\mathrm{c}}$ (bottom panel) $G_1(R,t) = \langle \hat{b}^{\dag}_R \hat{b}_0 \rangle_t - \langle \hat{b}^{\dag}_R \hat{b}_0 \rangle_0$ for a sudden global quench on the dissipation strength from $\gamma = 0$ to $\gamma > 0$ and on the two-body repulsive interaction strength $U$ from $U = U_{\mathrm{i}}$ to $U = U_{\mathrm{f}}$ such that the quantum system is initially confined in the SF-mean-field regime. The solid green line represents the correlation edge (CE) propagating linearly at the velocity $V_{\mathrm{CE}} = 2V_{\mathrm{g}}(k^*) = 2\mathrm{max}(\mathrm{d}\mathcal{E}_{k,\mathrm{f}}/\mathrm{d}k)$. The solid blue line represents the ballistic spreading of the series of local maxima and minima in the vicinity of the CE at the velocity $V_{\mathrm{m}} = 2V_{\varphi}(k^*) = 2\mathcal{E}_{k^*,\mathrm{f}}/k^*$. The considered post-quench and pre-quench dimensionless interaction parameters are $U_{\mathrm{f}}n(0)/J = 0.5$ and $U_{\mathrm{i}}n(0)/J = 2$ respectively. The dimensionless dissipation strength is given by $\gamma/J = 0.1$.}
\label{fig_g2_double_quench}
\end{figure}

\noindent
The latter is plotted on Fig.~\ref{fig_g2_double_quench} and we clearly see the effect of each quench individually. The quench on $U$ gives rise to a twofold linear structure characterized by a correlation edge (CE) propagating ballistically at twice the maximal group velocity. The second structure consists in a series of local maxima and minima spreading also ballistically in the vicinity of the CE at twice the phase velocity evaluated at the quasi-momentum for which the group velocity is maximal. This is reminiscent of the behavior of $G_2$ for a sudden global quench on the interaction strength $U$ for the isolated BH chain confined in the SF-mean-field regime. We can thus deduced from the previous findings that the algebraic decay in time of $G_2$, thus independent on the lattice site index $R$, corresponds to the signature of the quench on the dissipation strength $\gamma$. The latter effect is clearly visible in the background of Fig.~\ref{fig_g2_double_quench}. Similar statements are also valid for the $G_1$ ETCCF as depicted on Fig.~\ref{fig_g2_double_quench} and having the following expression, see also Eq.~\eqref{G1_th}, where $G_k(0) = \langle \hat{n}_0 \rangle_t$ is no longer a conserved quantity:
\begin{align}
G_1(R,t) =&~ \frac{1}{L}\sum_{k} \cos(kR) \left[G_k(t) - G_k(0)\right],
\end{align}
\noindent
while relying on the notations used to define the set of EoMs at Eq.~\eqref{EoMs}. 
\subsection{Quench spectroscopy of the double-quenched Bose-Hubbard chain}
\label{subsec:spectro}
In Refs.~\cite{despres_these,sanchezpalencia2019,villa2020}, the so-called quench spectroscopy (QS) method has been introduced theoretically whose predictions have been verified numerically using tensor-network-based techniques for one-dimensional (1D) quantum systems. The latter permits to determine the low-lying excitation spectrum or quasiparticle dispersion relation of quantum lattice models by analyzing their dynamical response to a sudden quench which can be either local, i.e. the perturbation is applied on a specific lattice site, or global, i.e. all the lattice sites are concerned. For instance, the QS has already been applied to disordered quantum lattice models~\cite{thomson2021,thomson2021paper2} and quantum spin models~\cite{menu2018,menu2023,bocini2024}. In what follows, we focus on the case of sudden global quenches. The main feature of this method corresponding also to the most restrictive condition is to consider a so-called weak quench meaning that the many-body ground state of the post-quench Hamiltonian has to be relatively closed to, i.e. strongly overlaps with, the initial state corresponding the ground state of the pre-quench Hamiltonian. In other words, the method holds for sudden global quenches confined in a same gapless or gapped phase and ideally in the same regime within this quantum phase. According to the QP theory used previously and detailed in Refs.~\cite{despres_these,despres2018}, equal-time connected correlation functions can generally be cast into a generic form. The latter reads as:
\begin{equation}
G(R,t) = \int_{\mathcal{B}}\mathrm{d}k\mathcal{S}_k \left[e^{i(kR+2\mathcal{E}_k t)} + e^{i(kR-2\mathcal{E}_k t)} \right],
\label{generic}
\end{equation}

\noindent
where $\mathcal{E}_{k}$ refers to the quasiparticle dispersion relation of the post-quench Hamiltonian. The so-called quench spectral function (QSF) denoted by
$S_k(\omega)$ is defined as the $1+1$ (space-time) Fourier transform of the equal-time connected correlation function $G(R,t)$ and reads as~\cite{villa2020}:
\begin{equation}
S_k(\omega) = \int_{0}^{L} \mathrm{d}R \int_{0}^{T}\mathrm{d}tG(R,t) e^{-i(kR + \omega t)},
\label{des}
\end{equation}

\noindent
where $T$ corresponds to the observation time and $L$ the system size or equivalently the total number of lattice sites. Inserting the generic form of
$G(R,t)$ at Eq.~\eqref{generic} in the definition of the QSF at Eq.~\eqref{des}, it yields in the thermodynamic limit, i.e. $L \rightarrow + \infty$, 
and in the limit of a large observation time, i.e. $T \rightarrow + \infty$, the following theoretical expression:
\begin{equation}
S_k(\omega) = \mathcal{S}_k \left[\delta(\omega + 2\mathcal{E}_{k}) + \delta(\omega - 2\mathcal{E}_{k}) \right].
\label{des2}
\end{equation}

\noindent
As expected, the previous form of the QSF clearly permits to determine the post-quench low-lying excitation spectrum $\mathcal{E}_{k}$. Indeed, from Eq.~\eqref{des2},
$S_k(\omega)$ represents twice the post-quench quasiparticle dispersion relation $2\mathcal{E}_k$. More precisely, the latter will display the two branches
$2E_{\mathbf{k}}$ and $-2E_{\mathbf{k}}$. The prefactor $2$ comes from the weak sudden global quench dynamics of the quantum lattice model being mediated by the spreading of quasiparticle pairs. The weight associated to each momentum-dependent energy is characterized by the amplitude function $\mathcal{S}_k$ depending on the quench parameters as well as the observable defining the equal-time correlation function $G(R,t)$. According to Eq.~\eqref{des2}, the QSF has $k$/$-k$ symmetry; the latter coming from the post-quench quasiparticle dispersion relation $\mathcal{E}_k$ and is $\omega$/$-\omega$ symmetric due to the parity of the Dirac delta function $\delta$. \\

For the case study of the BH chain submitted to a double quench, the ETCCF $G_2$ can be reformulated as follows:
\begin{equation}
G_2(R,t) = G_2^{U}(R,t) + G_2^{\gamma}(R,t) = G_2^{U}(R,t) + G_2^{\gamma}(0,t),
\label{G2_dq}
\end{equation}

\noindent
where $G_2^{U}(R,t)$ and $G_2^{\gamma}(R,t)$ refers to the individual contribution for $G_2$ due to the weak sudden global quench on the repulsive interaction strength $U$ and the dissipation rate $\gamma$ respectively. The first equality is valid according to the previous findings (see Fig.~\ref{fig_g2_double_quench} and the associated discussion) where both quantum quenches seem to be independent. The second one comes from the fact that the quench on $\gamma$ leads to a spatial-independent correlation spreading as depicted on Fig.~\ref{fig_g2_double_quench}. Indeed, from Ref.~\cite{despresnew} where an unique quench on the dissipation strength $\gamma$ is considered, $G_2^{\gamma}(R,t)$ reads as:

\begin{widetext}
\begin{subequations}
\begin{align}
G_2^{\gamma}(R,t) =&~ g_2^{\gamma}(R,t) - g_2^{\gamma}(R,0); \\
G_2^{\gamma}(R,t) =&~ n(t)^2 + \frac{2n(t)}{L}\sum_{k \neq 0} \cos(kR)\left[G_k(t) + \operatorname{Re}(F_k(t))\right] - n(0)^2 - \frac{2n(0)}{L}\sum_{k \neq 0} \cos(kR)\left[G_k(0) + \operatorname{Re}(F_k(0))\right],
\label{G2_gamma}
\end{align}
\end{subequations}
\end{widetext}

\noindent
where $g_2(R,t)$ is defined at Eq.~\eqref{eq_g2}. For the BH chain in the SF-mean-field, the first excitations having a momentum $k = 0^+$ are the relevant ones. This implies that the associated wavelength $\lambda$
is much higher than the lattice spacing $a$, i.e. $\lambda \gg a$. Hence, it follows that the discretization of the lattice is irrelevant and thus that the contribution $G_2^{\gamma}(R,t)$ is spatially independent, i.e. $G_2^{\gamma}(R,t) = G_2^{\gamma}(0,t)$. This can be verified straightforwardly by injecting $k = 0^+$ in Eq.~\eqref{G2_dq}. Finally, it immediately follows from Eq.~\eqref{G2_dq} that: 
\begin{equation}
S_k(\omega) = S_k^U(\omega) + S_{k = 0}^{\gamma}(\omega).
\label{skw_G2}
\end{equation}

\begin{figure}[h!]
\centering
\includegraphics[scale = 0.43]{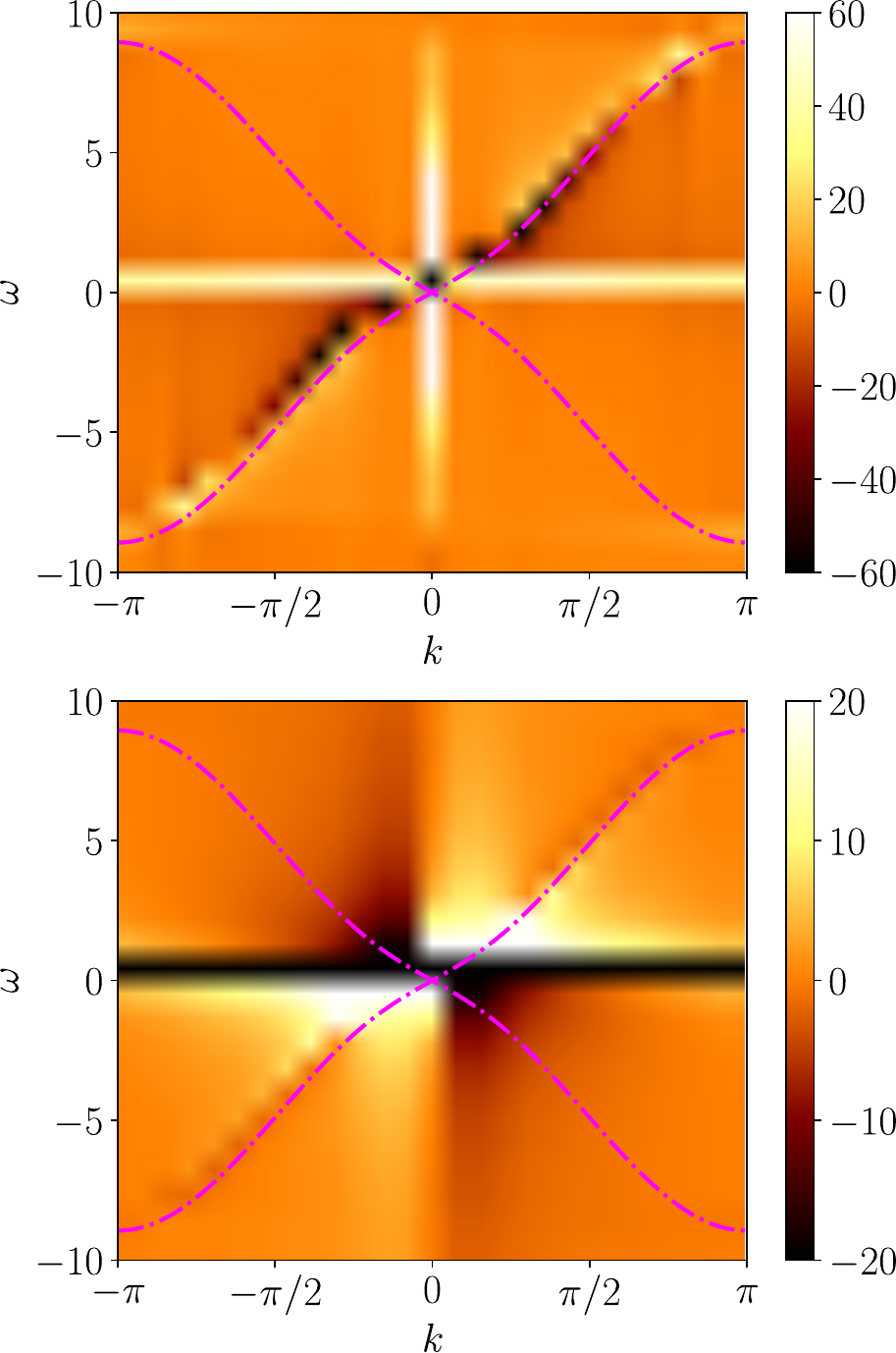}
\caption{Quench spectral function $S_k(\omega)$ associated to (top panel) $G_2(R,t) = \langle \hat{n}_{R}(t) \hat{n}_{0}(t) \rangle_{\mathrm{c}}$ the equal-time connected density-density correlation function (bottom panel) $G_1(R,t) = \langle \hat{b}^{\dag}_R \hat{b}_0 \rangle_t - \langle \hat{b}^{\dag}_R \hat{b}_0 \rangle_0$ the equal-time connected one-body correlation function. The double quench here comprises a weak sudden global quench on the dissipation strength from $\gamma = 0$ to $\gamma > 0$ as well as one performed on the two-body repulsive interaction strength $U$ from $U = U_{\mathrm{i}}$ to $U = U_{\mathrm{f}}$. The dashed-dotted magenta lines represent twice and minus twice the post-quench quasiparticle dispersion relation associated to the isolated BH chain confined in the SF-mean-field regime, i.e. $2\mathcal{E}_{k,\mathrm{f}}$ and $-2\mathcal{E}_{k,\mathrm{f}}$ respectively, see Eq.~\eqref{twice_qdr}. Note that half of the two low-lying excitation spectra are missing resulting in a loss of the $k/-k$ and $\omega/-\omega$ symmetries. Indeed, the latter would be characterized by a negative group velocity and only the correlation pattern with $R,t \geq 0$ is considered here. To recover the second half, the space-time region of the correlation pattern with $R \leq 0$ and $t \geq 0$ has to be considered. The considered post-quench and pre-quench dimensionless interaction parameters are $U_{\mathrm{f}}n(0)/J = 0.5$ and $U_{\mathrm{i}}n(0)/J = 2$ respectively. The dimensionless dissipation strength is given by $\gamma/J = 0.1$.}
\label{fig_G2_G1_qsf}
\end{figure}

\noindent
Therefore, the consequence of the quench on $\gamma$ for the quench spectral function is minimal according to Eq.~\eqref{skw_G2}. Indeed, the latter will increase the amplitude of $S_k(\omega)$ for the quasi-momentum $k = 0$ for any energy $\omega$. This effect is shown on Fig.~\ref{fig_G2_G1_qsf}. On Fig.~\ref{fig_G2_G1_qsf}, the QSFs associated to the $G_2$ and $G_1$ ETCCF are displayed and are benchmarked with twice the post-quench quasiparticle dispersion relation associated to the isolated BH chain confined in the SF-mean-field regime, i.e. $2\mathcal{E}_k$, having the following expression:
\begin{equation}
2\mathcal{E}_{k,\mathrm{f}} = 4\sqrt{ 2J\sin^2(k/2)\left( 2J\sin^2(k/2) + U_{\mathrm{f}}\bar{n} \right)},
\label{twice_qdr}
\end{equation} 

\noindent
where $\bar{n} = N(0)/L$ refers here to the initial filling of the lattice chain with $N(0)$ denoting the initial total number of bosonic particles on the lattice, see also Eq.~\eqref{eq_spectrum}. Note that for $G_1$,
the spatial dependence of the contribution due to the quench on the dissipations, i.e. $G_1^{\gamma}(R,t)$, becomes slightly relevant as suggested on Fig.~\ref{fig_g2_double_quench}. This results in non-zero amplitudes for the associated QSF for a momentum $k$ in the limit $k \rightarrow 0$ and for any energy $\omega$. The latter is consistent with the fact that the first excitations
have a momentum close to zero. \\

\begin{figure}[h!]
\centering
\includegraphics[scale = 0.37]{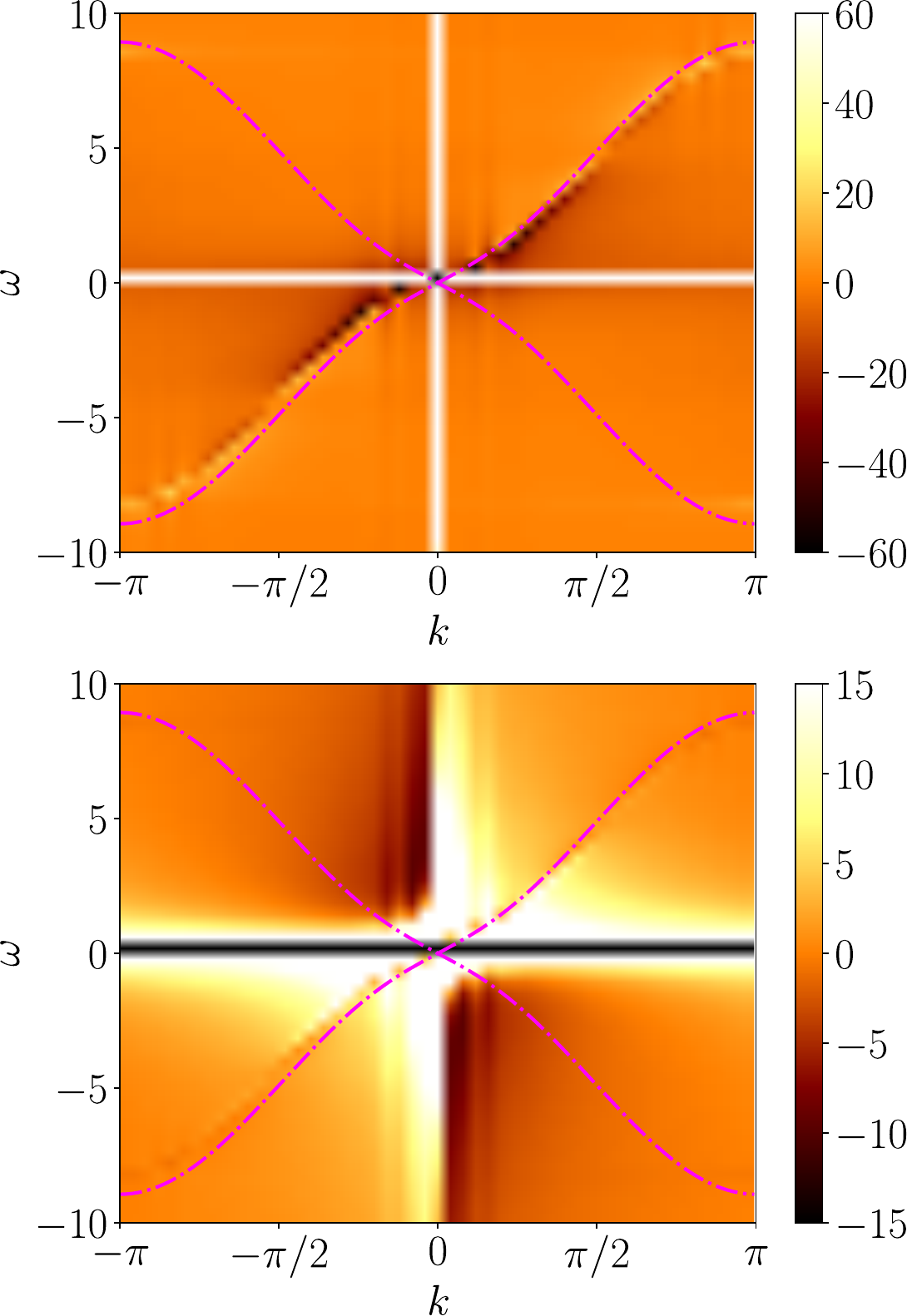}
\caption{Quench spectral function $S_k(\omega)$ associated to (top panel) $G_2(R,t) = \langle \hat{n}_{R}(t) \hat{n}_{0}(t) \rangle_{\mathrm{c}}$ the equal-time connected density-density correlation function (bottom panel) $G_1(R,t) = \langle \hat{b}^{\dag}_R \hat{b}_0 \rangle_t - \langle \hat{b}^{\dag}_R \hat{b}_0 \rangle_0$ the equal-time connected one-body correlation function. A weak sudden double quench is performed on the dissipation strength from $\gamma = 0$ to $\gamma > 0$ as well as on the two-body repulsive interaction strength $U$ from $U = U_{\mathrm{i}}$ to $U = U_{\mathrm{f}}$. The dashed-dotted magenta lines represent twice and minus twice the post-quench quasiparticle dispersion relation associated to the isolated BH chain confined in the SF-mean-field regime, i.e. $2\mathcal{E}_{k,\mathrm{f}}$ and $-2\mathcal{E}_{k,\mathrm{f}}$ respectively, see Eq.~\eqref{twice_qdr}. The considered post-quench and pre-quench dimensionless interaction parameters are $U_{\mathrm{f}}n(0)/J = 0.5$ and $U_{\mathrm{i}}n(0)/J = 2$ respectively. The dimensionless dissipation strength is given by $\gamma/J = 1$.}
\label{fig_G2_G1_qsf_large_gamma}
\end{figure}

We move on to the case of strong dissipations implying large $\gamma$. The aim of this investigation is to certify the validity of the QS approach for strong loss processes. 
On Fig.~\ref{fig_G2_G1_qsf_large_gamma}, we consider a dissipation strength ten times higher than previously for the double-quenched BH chain confined in the SF phase. We notice that the QSF is well benchmarked by the twice (or minus twice) $\mathcal{E}_{k, \mathrm{f}}$ the theoretical quasiparticle dispersion relation associated to the post-quench Hamiltonian of the BH chain confined in the SF phase. Consequently, this permits to state that the QS method can still be applied in the framework of strong loss processes. \\

\begin{figure}[h!]
\centering
\includegraphics[scale = 0.35]{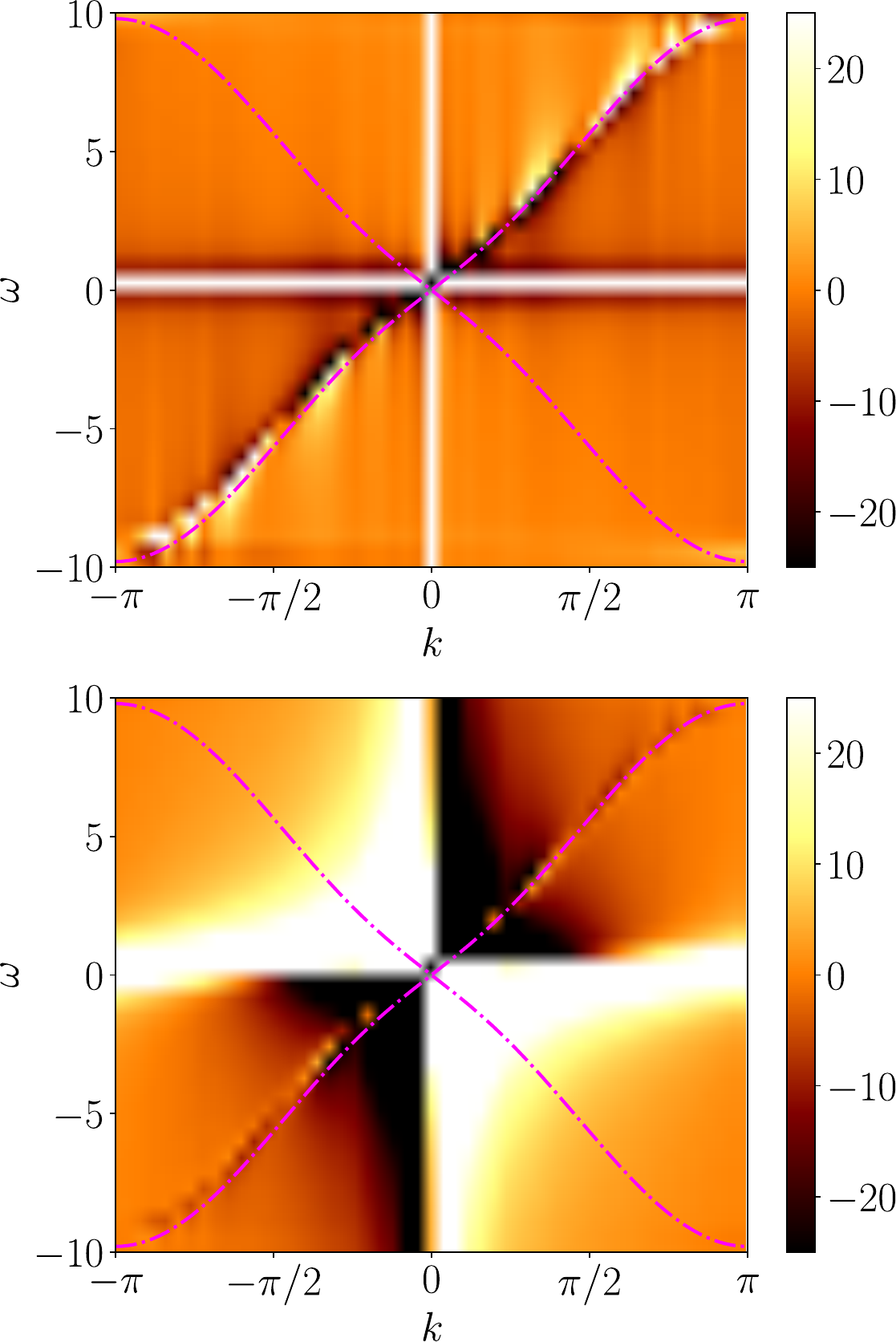}
\caption{Quench spectral function $S_k(\omega)$ associated to (top panel) $G_2(R,t) = \langle \hat{n}_{R}(t) \hat{n}_{0}(t) \rangle_{\mathrm{c}}$ the equal-time connected density-density correlation function (bottom panel) $G_1(R,t) = \langle \hat{b}^{\dag}_R \hat{b}_0 \rangle_t - \langle \hat{b}^{\dag}_R \hat{b}_0 \rangle_0$ the equal-time connected one-body correlation function. A weak sudden double quench is performed on the dissipation strength from $\gamma = 0$ to $\gamma > 0$ as well as on the two-body repulsive interaction strength $U$ from $U = U_{\mathrm{i}}$ to a relatively large post-quench interaction strength $U = U_{\mathrm{f}}$. The dashed-dotted magenta lines represent twice and minus twice the post-quench quasiparticle dispersion relation associated to the isolated BH chain confined in the SF-mean-field regime, i.e. $2\mathcal{E}_{k,\mathrm{f}}$ and $-2\mathcal{E}_{k,\mathrm{f}}$ respectively, see Eq.~\eqref{twice_qdr}. The considered post-quench and pre-quench dimensionless interaction parameters are $U_{\mathrm{f}}n(0)/J = 1$ and $U_{\mathrm{i}}n(0)/J = 2$ respectively. The dimensionless dissipation strength is given by $\gamma/J = 0.1$.}
\label{fig_G2_G1_qsf_large_post_U}
\end{figure}

We turn to the case of relatively large post-quench interaction strength $U_{\mathrm{f}}$ while considering small dissipations, i.e. $\gamma \ll 1$. According to the mean-field condition given by $\bar{n} \gg U/J$, for large enough $U_{\mathrm{f}}$, the 1D BH model at equilibrium is no longer confined in the mean-field regime of the SF phase but rather in the strongly-correlated regime. In what follows, we consider the specific case where $J = U_{\mathrm{f}} = \bar{n} = N(0)/L$. However, from Ref.~\cite{despres2019}, it has been shown that the quasiparticle dispersion relation $\mathcal{E}_{k}$ for the BH chain confined in the SF-mean-field regime defined at Eq.~\eqref{twice_qdr} remains accurate far beyond its validity domain. Hence, even at relatively large post-quench interaction strength $U_{\mathrm{f}}$, we should be able to rely on the QS approach in order to get an accurate description of the low-lying excitation spectrum of the 1D BH model confined in the SF-strongly-correlated regime. On Fig.~\ref{fig_G2_G1_qsf_large_post_U}, the QSF associated to the density fluctuations characterized via the ETCCF $G_2$ and to the phase fluctuations via $G_1$ are compared to the theoretical expression of the quasiparticle dispersion relation $\mathcal{E}_{k,\mathrm{f}}$ initially valid in the SF-mean-field regime. As shown on both panels, we find a very good agreement between the low-lying excitation spectra found from the QSFs and $\mathcal{E}_{k,\mathrm{f}}$. \\

We stress that all the previous findings do not apply for sudden global quenches confined in the MI phase. Indeed, to generate the corresponding low-lying
excitations, it requires to consider a subspace of the full Hilbert space where the $U(1)$ particle-number symmetry is conserved during the dynamics. Hence, in the presence of loss processes, the latter is broken and the quantum system remains in the SF phase independently of the strength of the repulsive interactions due to the incommensurate filling of the lattice chain.

\section{Experimental feasibility of the quench spectroscopy method}
\label{feasibility}
The QS presents several difficulties of applicability and limits from an experimental point of view. Indeed, in the context of the dissipative BH chain, the main limitations are the following:\\

$\tiny \bullet$ The restrictive choice for the observables in order to compute the associated connected equal-time correlation function.\\

$\tiny \bullet$  The different sources of noise including the shot and technical noise.\\ 

$\tiny \bullet$  The difficulty for the imaging due to the coherence of the time-evolved quantum state when considering the gapless SF phase contrary to the gapped MI phase. \\

$\tiny \bullet$ The resolution associated to the experimental data to describe accurately the causal region of the space-time correlations. The latter region is primordial to recover the full information regarding the low-lying excitation spectrum of the quantum lattice model. Indeed, reliable experimental data on the correlation edge, which is already not an easy task for experimentalists, is not sufficient. The latter contains only information on the quasiparticle pair propagating with the fastest group velocity.\\

$\tiny \bullet$  From a more general viewpoint, a main limit of the QS method is to be valid and reliable for weak global quenches. This "quench weakness" implies difficulties for experimentalists. To illustrate the previous difficulty, a strong global quench confined in the MI phase is much easier to handle compared to a weak quench in the same phase. Indeed, for the imaging and more precisely for the mobility of the bosonic particles, a strong global quench is necessary. By "strong quench", we mean here that the initial or pre-quench dimensionless interaction parameter is significantly higher (or lower) than the final or post-quench dimensionless interaction parameter, i.e. $(U/J)_{\mathrm{i}} \gg (U/J)_{\mathrm{f}}$.\\

$\tiny \bullet$ Another limitation of the method lies in the kind of loss processes which is considered. It is entirely possible to imagine loss processes leading to a non-trivial, i.e. strong, spatial dependence for the space-time pattern of the ETCCF. \\

The latter QS method to probe low-lying excitation spectra can be extended to dissipative many-body quantum fermionic and spin lattice chains, to a higher dimensionality of the lattice and to long-range interactions. Most importantly, this technique can be applied to experimental measurements obtained using quantum simulators based on ultra-cold atoms to engineer bosonic lattice models, neutral Rydberg atoms or trapped ions for the simulation of fermionic and spin lattice models. To conclude, QS can be seen as an alternative approach to the standard experimental techniques to measure excitation spectra such as the angle-resolved photoemission spectroscopy (ARPES) ~\cite{stewart2008} or Bragg spectroscopy \cite{rey2005}.

\section{Applicability of the quench spectroscopy to a non-Hermitian quantum lattice model}
\label{applicability}
We focus here on the case study of the $s = 1/2$ non-Hermitian transverse-field Ising model (TFIM) on a one-dimensional chain of $L$ lattice sites whose lattice spacing is still fixed to unity, i.e. $a = 1$; for simplicity, we still consider $\hbar = 1$. The corresponding Hamiltonian $\hat{H}$ reads:
\begin{equation}
\hat{H} =  J \sum_R \hat{S}^x_R \hat{S}^x_{R+1} - (h + i\gamma) \sum_R \hat{S}^z_R,
\label{H_ising_chain}
\end{equation}
\noindent
where $\hat{S}^{\alpha}_R$ corresponds to the $s = 1/2$ spin operator acting on the lattice site $R$ along the $\alpha \in \{x,z\}$ axis. $J > 0$ refers to the antiferromagnetic spin exchange coupling between nearest neighbors spins, $h > 0$ is the constant and homogeneous transverse magnetic field and $\gamma $ denotes the dissipative strength representing the local dephasing noise. In what follows, we consider $\gamma > 0$ as well as the paramagnetic phase (also called $z$ polarized phase in the literature) where the spins are polarized along $z$ axis, i.e. are aligned with the magnetic field, for $h \gg J$. In Ref.~\cite{despres2024}, the Authors studied the quasiparticle dispersion relation as well as the quench dynamics of this non-Hermitian spin model in the paramagnetic phase. Regarding the low-lying excitation spectrum, they relied on the following Holstein-Primakoff (HP) transformation:
\begin{align}
& \hat{S}^x_R = \frac{\hat{a}_R + \hat{a}^{\dag}_R}{2},~\hat{S}^{y}_R = \frac{\hat{a}_R - \hat{a}^{\dag}_R}{2i},~\hat{S}^z_R = \frac{1}{2} - \hat{a}^{\dag}_R \hat{a}_R,
\end{align}

\noindent
where $\hat{a}_R$ ($\hat{a}^{\dag}_R$) refers to the annihilation (creation) bosonic operator acting on the lattice site $R$. The latter obey canonical commutation relations. This HP transformation implies to consider a regime in which the bosonic local occupation number remains small, i.e. $\langle \hat{a}^{\dag}_R \hat{a}_R \rangle \ll 1$, which is the case for example in the paramagnetic phase of the TFIM at equilibrium. This transformation permits to express $\hat{H}$ at Eq.~\eqref{H_ising_chain} in the generic quadratic Bose form discussed at Eq.~\eqref{eq_bhm_quadratic} with $\mathcal{A}_k = h + i\gamma + \mathcal{B}_k$ and $\mathcal{B}_k = (J/2) \cos(k)$. Note that here the value $k = 0$ is included in the summation over the momentum. \\

To investigate the quench dynamics of the non-Hermitian TFIM, the Authors of Ref.~\cite{despres2024} derived the EoM associated to the quadratic bosonic correlator 
$G_k(t) = \langle \hat{a}^{\dag}_k \hat{a}_k \rangle_t$ and $F_k(t) = \langle \hat{a}_k \hat{a}_{-k} \rangle_t$ where $\langle ... \rangle_t = \langle \Psi(t) | ... | \Psi(t) \rangle$. The time-evolved many-body quantum state $\ket{\Psi(t)}$ is defined as:
\begin{equation}
\ket{\Psi(t)} = \frac{e^{-i\hat{H}t}\ket{\Psi_0}}{||e^{-i\hat{H}t}\ket{\Psi_0}||},
\label{psi_t}
\end{equation}
\noindent
where $\ket{\Psi_0} = \ket{\Psi(0)} = \ket{\mathrm{GS}(\hat{H}(\gamma = 0))}$ denotes the initial many-body quantum state corresponding to the ground state of the Hermitian version of the Hamiltonian $\hat{H}$ implying $\gamma = 0$. Using the latter expression, the EoM associated to the time-dependent expectation value of any observable $\hat{A}$ denoted by $G_A(t) = \langle \Psi(t)|\hat{A}|\Psi(t)\rangle = \langle \hat{A} \rangle_t$ reads as:
\begin{align}
& \frac{\mathrm{d}}{\mathrm{d}t} G_A(t) = i \langle \hat{H}^{\dag}\hat{A}-\hat{A}\hat{H}\rangle_t + i\langle \hat{H}-\hat{H}^{\dag}\rangle_t G_A(t).
\end{align}
\noindent
Then, by replacing $\hat{A}$ by $\hat{a}^{\dag}_k \hat{a}_k$ for the correlator $G_k(t)$ and by $\hat{a}_k \hat{a}_{-k}$ for $F_k(t)$, the following set of non-linear coupled differential equations has been found: 
\begin{widetext}
\begin{subequations}
\label{diff_eq_G_k_t}
\begin{align}
\frac{\mathrm{d}}{\mathrm{d}t}G_k(t) =&~ -2\mathcal{B}_k\operatorname{Im}(F_k(t)) + 2\operatorname{Im}(\mathcal{A}_k)\left(|F_k(t)|^2 + G_k(t) + G_k(t)^2 \right);\\
\frac{\mathrm{d}}{\mathrm{d}t}F_k(t) =&~ 4\operatorname{Im}(\mathcal{A}_k)F_k(t)G_k(t) - 2i\mathcal{A}_kF_k(t) -i\mathcal{B}_k \left[1+2G_k(t)\right],
\end{align}
\end{subequations}
\end{widetext}

\begin{figure}[h!]
\centering
\includegraphics[scale = 0.49]{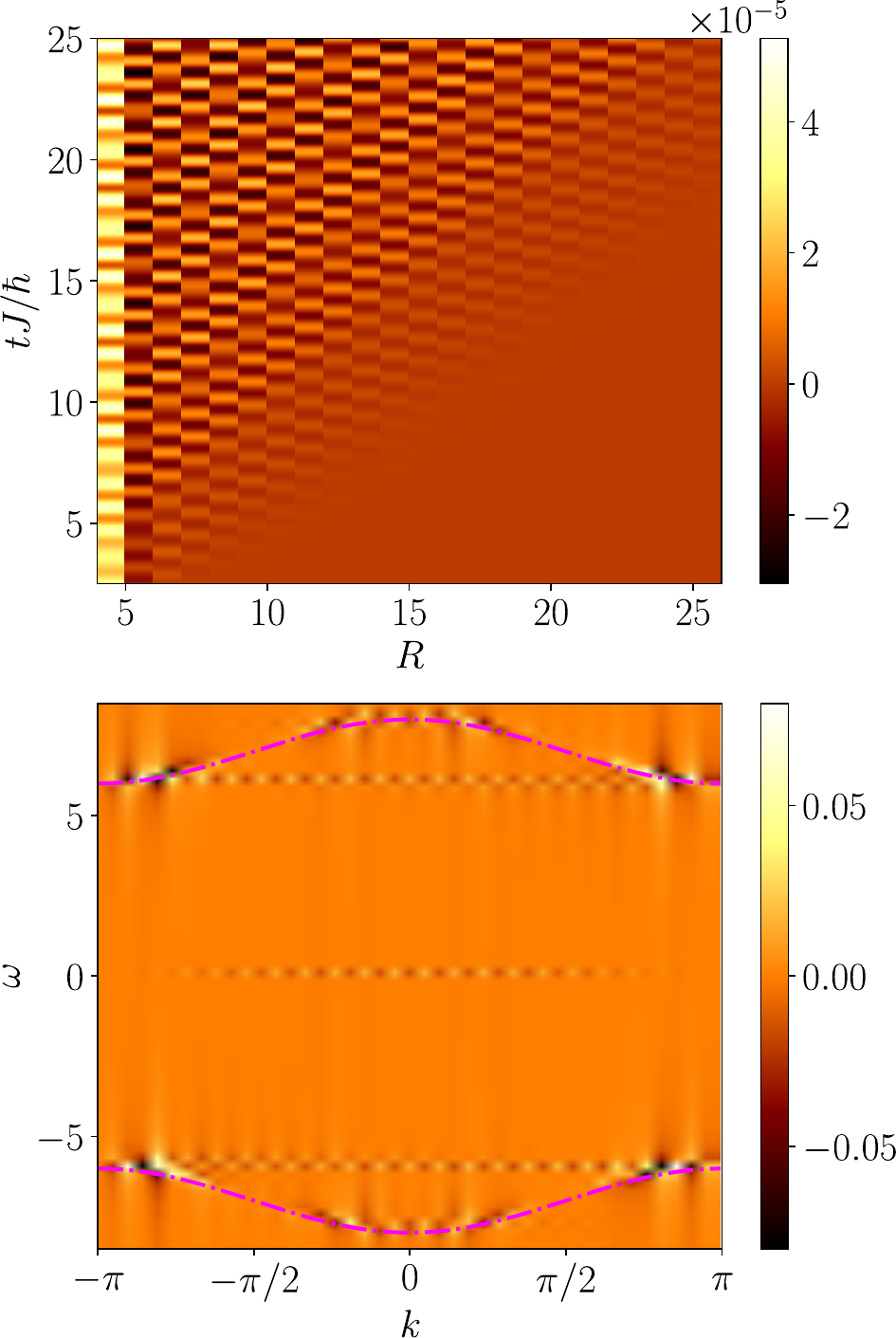}
\caption{(Top panel) $G_1(R,t) = \langle \hat{a}^{\dag}_R \hat{a}_0 \rangle_t$ the equal-time one-body correlation function (bottom panel) the associated quench spectral function $S_k(\omega)$ for a weak sudden global quench on $\gamma$ from $\gamma = 0$ to $\gamma > 0$ for the non-Hermitian 1D TFIM confined in the paramagnetic ($z$ polarized) phase. The dashed-dotted magenta lines represent twice and minus twice the real part of the momentum-dependent function $\mathcal{A}_k$, i.e. $2\operatorname{Re}(\mathcal{A}_k)$ and $-2\operatorname{Re}(\mathcal{A}_k)$ respectively. For clarity, the space-time correlation pattern of $G_1$ for the calculation of $S_k(\omega)$ 
is plotted for $R \geq 0$ and $R \leq 0$ and $t \geq 0$ to recover the second half of the effective quasiparticle dispersion relation. The dimensionless transverse-field and dissipation parameters are given by $h/J = 3.5$ and
$\gamma/J = 0.02$ respectively.}
\label{fig_G1_qsf}
\end{figure}

\noindent
whose initial values are given by:
\begin{subequations}
\begin{align}
& F_k(0) = \cosh(\alpha_{k})\sinh(\alpha_{k}); \\
& G_k(0) = \sinh^2(\alpha_{k}), \\
& \alpha_k = \frac{1}{2}\mathrm{arctanh} \left(-\frac{\mathcal{B}_k}{\mathcal{A}_{k}(\gamma = 0)}\right).
\end{align}
\end{subequations}

\noindent
Using the set of EoMs defined at Eq.~\eqref{diff_eq_G_k_t}, the correlation spreading induced by the quench on $\gamma$ is investigated in Ref.~\cite{despres2024} by analyzing the equal-time one-body correlation function $G_1(R,t)$ which reads as:
\begin{align}
G_1(R,t) = \langle \hat{a}^{\dag}_R \hat{a}_0 \rangle_t = \frac{1}{L} \sum_k \cos(kR) G_k(t),
\label{G_R_t_G_k_t}
\end{align}

\noindent
whose space-time pattern deduced from the previous set of EoMs is qualitatively very well reproduced by the following theoretical guess:
\begin{align}
& G_1(R,t) = \frac{1}{L} \sum_k \mathcal{F}_k^{(1)} \cos(kR) \cos[2\operatorname{Re}(\mathcal{A}_k)t]e^{2\operatorname{Im}(\mathcal{A}_k)t},
\label{G_R_t_th_guess}
\end{align}

\noindent
where $\operatorname{Re}(\mathcal{A}_k) = h + (J/2)\cos(k)$ and  $\operatorname{Im}(\mathcal{A}_k) = \gamma$. It immediately follows that $G_1(R,t)$ can be cast into the generic form previously studied. Indeed, after performing some algebra and by considering the thermodynamic limit, i.e. $L \rightarrow + \infty$, we get:
\begin{equation}
G_1(R,t) \sim \int_{\mathcal{B}}\mathrm{d}k \mathcal{S}_k^{(1)} \left\{e^{i[kR+2\operatorname{Re}(\mathcal{A}_k)t]} + e^{i[kR-2\operatorname{Re}(\mathcal{A}_k)t]}\right\}.
\label{G1_nh_tfim}
\end{equation}

\noindent
From Eq.~\eqref{G1_nh_tfim} and by performing the Fourier transform, we can easily deduce that only (twice) the real part of the complex low-lying excitation spectrum of the non-Hermitian 1D TFIM will be unveiled by the QSF. Indeed, in the paramagnetic phase, the first band of the spectrum is given by \cite{despres2024}:
$\mathcal{E}_k = \mathrm{sgn}(\operatorname{Re}(\mathcal{A}_k))\sqrt{\mathcal{A}_k^2 - \mathcal{B}_k^2}$. Deep in the latter phase where $h \gg J$, we get:
$\mathcal{E}_k = \mathcal{A}_k$ hence $\operatorname{Re}(\mathcal{E}_k) = \operatorname{Re}(\mathcal{A}_k)$. This statement is depicted on Fig.~\ref{fig_G1_qsf} where the QSF probes twice the real part of the quasiparticle dispersion relation $2\operatorname{Re}(\mathcal{E}_k) = 2\operatorname{Re}(\mathcal{A}_k)$. The latter investigation permits to certify the applicability and reliability of the QS method for non-Hermitian quantum lattice models and can be straightforwardly extended to non-Hermitian quantum systems subjected to long-range interactions and for a higher dimensionality of the lattice. \\

Note that the QP approach is not adapted to study the quench dynamics of isolated non-Hermitian quantum lattice models. Indeed, the reason is twofold. On one hand, the Hamiltonian $\hat{H}$ being non-Hermitian
i.e. $\hat{H} \neq \hat{H}^{\dag}$, the time-dependent operators will involve both Hamiltonians. For instance, a time-dependent operator $\hat{O}$ can be expressed as
$\hat{O}(t) = e^{i\hat{H}^{\dag}t} \hat{O} e^{-i\hat{H}t}$ using the Heisenberg picture. Then, for our case study, the Hamiltonian of the non-Hermitian transverse-field Ising chain in the paramagnetic phase can be diagonalized using a non-standard (fermionic or bosonic) Bogolyubov transformation \cite{despres2024}. For instance, for bosonic Bogolyubov operators, $\hat{H}$ can be diagonalized in terms of $\hat{b}$ and $\hat{\bar{b}}$ where $\hat{\bar{b}} \neq \hat{b}^{\dag}$. This implies to deduce as well as to work analytically with four operators namely $\hat{b}(t)$, $\hat{\bar{b}}(t)$, $\hat{b}^{\dag}(t)$ and $\hat{\bar{b}}^{\dag}(t)$. On the other hand, since a non-Hermitian Hamiltonian $\hat{H}$ is considered leading to the a non-conservation of the norm of the time-evolved quantum state, see Eq.~\eqref{psi_t}, the latter has to be calculated analytically and involving the four operators defined previously, i.e. $|| \ket{\Psi(t)} ||^2 = \langle \Psi_0 | e^{i(\hat{H}^{\dag} - \hat{H})t} | \Psi_0 \rangle$. 

\section{Correlation spreading in the transverse-field Ising chain}
\label{sec:srti}
In whats follows, we consider the $s = 1/2$ short-range interacting transverse-field Ising (TFI) chain. As shown below, its Hamiltonian can be reformulated in a quadratic bosonic or fermionic form, i.e. involving a bosonic or fermionic Bogolyubov-de-Gennes Hamiltonian respectively. The correlation spreading of the latter model is investigated; one main goal being to attest of the universality of the twofold linear structure for quadratic bosonic and fermionic quantum systems. 
To do so, we consider the bosonic and fermionic reformulation of the model and calculate analytically an ETCCF to study its quench dynamics. This will be performed using the EoM and QP theoretical approaches. 

\subsection{Bosonic reformulation}
\label{subsec:boson}
The Hermitian Hamiltonian of the TFI chain is already discussed at Section~\ref{applicability} for $\gamma = 0$, see Eq.~\eqref{H_ising_chain}. To unravel its quench dynamics features, a sudden global 
quench on the homogeneous and constant external field $h$ is considered. The initial (pre-quench) and final (post-quench) values are chosen such that the spin model is confined within the paramagnetic phase; the latter implying a dimensionless interaction parameter $h/J \gg 1$. We first consider the bosonic reformulation of the TFI chain whose corresponding low-lying excitations can be seen as bosonic Bogolyubov ones. 

\paragraph{Bosonic Bogolyubov theory} To unveil the quadratic Bose form, we rely on the Holstein-Primakoff (HP) transformation suitable for the paramagnetic ($z$-polarized) phase given by: 
\begin{align}
& \hat{S}^+_R = \hat{a}_R,~~~~~~\hat{S}^{-}_R = \hat{a}^{\dag}_R,~~~~~~\hat{S}^z_R = \frac{1}{2} - \hat{a}^{\dag}_R \hat{a}_R.
\label{hp}
\end{align}
The second order approximation in terms of $\hat{a}^{(\dag)}$ is valid provided that $\langle \hat{a}^{\dag}_R \hat{a}_R \rangle_t \ll 1$, $\forall (R,t) \in \mathbb{N} \times \mathbb{R}$. This property is verified within our case study since a sudden global quench confined within the paramagnetic phase is considered. It yields the quadratic Bose form:
\begin{equation}
\hat{H} = \frac{1}{2} \sum_{k}A_{k}\left(\hat{a}^{\dag}_{k} \hat{a}_{k}+\hat{a}_{-k}\hat{a}^{\dag}_{-k}\right)+B_{k} \left(\hat{a}^{\dag}_{k} \hat{a}^{\dag}_{-k}+\hat{a}_{k} \hat{a}_{-k} \right),
\label{eq_bhm_quadratic}
\end{equation}
\noindent
with $A_{k} = h + (J/2)\cos(k)$ and $B_{k} = (J/2)\cos(k)$. We then use the following Bogolyubov transformation while introducing bosonic Bogolyubov operators denoted by $\hat{b}^{(\dag)}$:
\begin{equation}
\hat{a}_k = u_k \hat{b}_{k} + v_k \hat{b}^{\dag}_{-k},~~~~~~\hat{a}^{\dag}_k = u_k \hat{b}^{\dag}_{k} + v_k \hat{b}_{-k},
\end{equation}
where $u_k$, $v_k$ are real and even momentum-dependent functions having the following expression: 
\begin{subequations}
\begin{align}
& u_k = \cosh\left(\theta_k \right); \\
& v_k = \sinh\left(\theta_k\right), \\
& 2\theta_k = \mathrm{arctanh}\left(-\frac{B_k}{A_k}\right).
\end{align} 
\end{subequations}
Up to a shift in energy, i.e. a constant, the Hamiltonian $\hat{H}$ is diagonalized and reads as: 
\begin{equation}
\hat{H} = \sum_k E_k \hat{b}^{\dag}_k \hat{b}_k,~~E_k = \mathrm{sgn}(A_k) \sqrt{A_k^2 -B_k^2}. 
\label{diag_boson}
\end{equation} 
\label{sec:benchmark}

\begin{figure*}[ht]
\centering
\includegraphics[scale = 0.445]{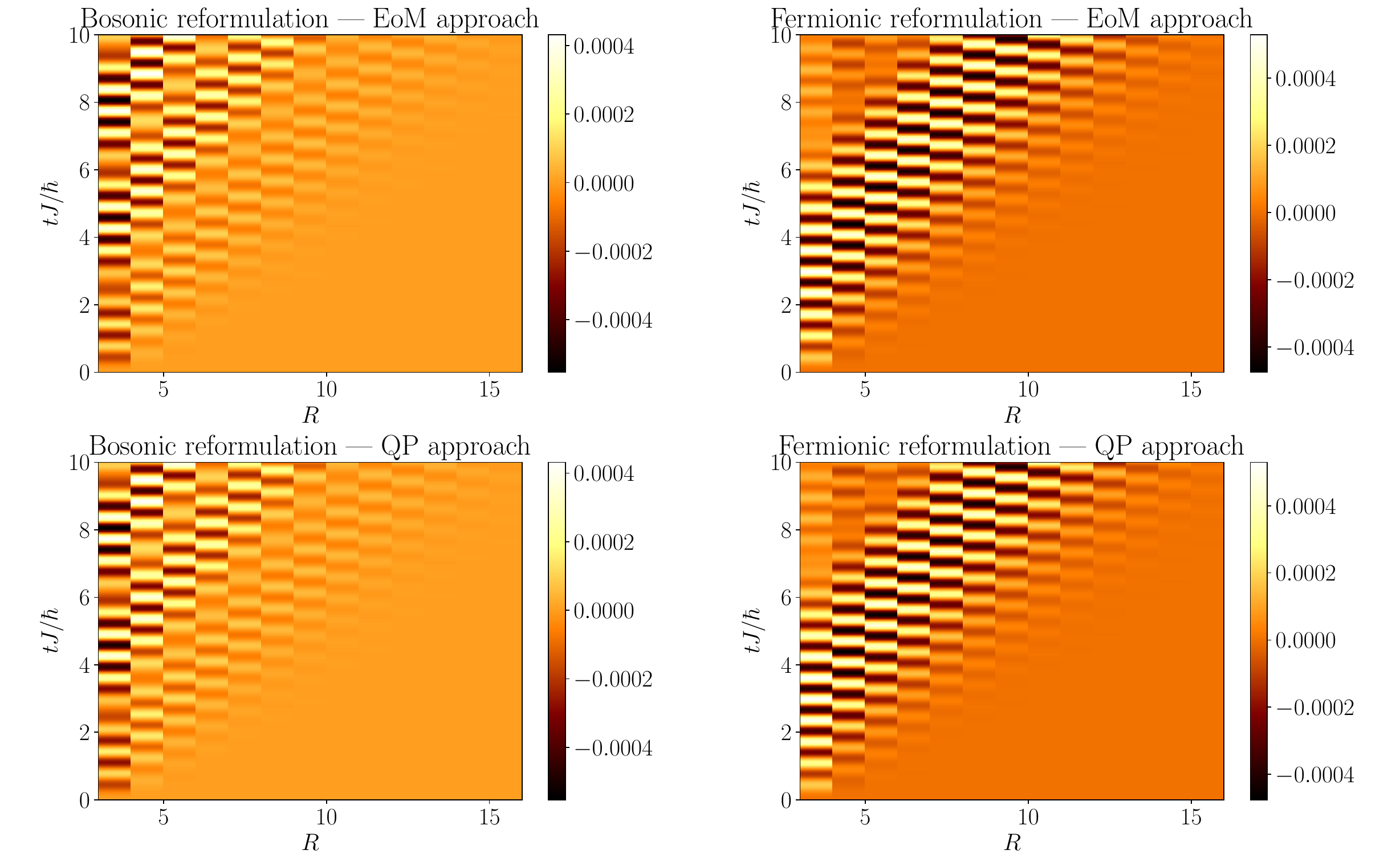}
\caption{Sudden global quench dynamics of the TFI chain confined within the paramagnetic phase: $G_1(R,t)$ the equal-time connected one-body correlation function is plotted as a function of the lattice site index $R$ and the dimensionless time $tJ/\hbar$ via a bosonic reformulation leading to $G_1(R,t) = \langle \hat{a}_R^{\dag}(t)\hat{a}_0(t) \rangle_c$ while benchmarking the EoM approach using Eq.~\eqref{EoMs_boson} (top left panel) with the QP approach using Eq.~\eqref{qp_boson} (bottom left panel). The fermionic reformulation is also considered where the ETCCF can be expressed as follows $G_1(R,t) = \langle \hat{c}_R^{\dag}(t)\hat{c}_0(t) \rangle_c$ while again benchmarking the EoM approach using Eq.~\eqref{EoMs_fermion} (top right panel) with the QP approach using Eq.~\eqref{qp_fermion} (bottom right panel). For all the different panels, the parameters are: $J = 1$, $h_{\mathrm{i}} = 8$, $h_{\mathrm{f}} = 5$. A very good agreement is found between the four cases. Note that here the expected twofold ballistic correlation spreading is characterized by a negative velocity for the spreading of the series of local extrema. The latter property can be easily explained by the shape of the quasiparticle dispersion relation of the TFI chain confined in the paramagnetic phase where the phase velocity $V_{\varphi}(k)$ is negative when the group velocity $V_{\mathrm{g}}(k)$ is positive.}
\label{srti}
\end{figure*}

\paragraph{EoM approach} We first consider the EoM approach. We can notice that the bosonic reformulation of the TFI chain confined in the paramagnetic phase has similar features than the Bose-Hubbard chain in the SF-mean-field regime presented at Section~\ref{sec:benchmark}.
It immediately follows the same set of EoMs given by: 
\begin{subequations}
\label{EoMs_boson}
\begin{align}
\frac{\mathrm{d}}{\mathrm{d}t} G_k(t) =&~ -2B_{k,\mathrm{f}}\operatorname{Im}[F_k(t)],\\
\frac{\mathrm{d}}{\mathrm{d}t} F_k(t) =&~ -2iA_{k,\mathrm{f}}F_k(t) - iB_{k,\mathrm{f}}[1+2G_k(t)].
\end{align}
\end{subequations}
\noindent
As a reminder, the quadratic correlators are defined as $G_{k}(t) = \langle \hat{a}^{\dag}_{k} \hat{a}_{k} \rangle_t = \langle \hat{n}_{k} \rangle_t$ and $F_{k}(t) = \langle \hat{b}_{-k} \hat{b}_{k} \rangle_t$. 
The coefficients $A_{k,\alpha}$ and $B_{k,\alpha}$ where $\alpha \in \{\mathrm{i},\mathrm{f}\}$ denote the initial (index i) and the final (index f) momentum-dependent functions valid for the pre-quench and post-quench TFI Hamiltonians respectively.
The latter will thus depend on $h_{\mathrm{i}}$ or $h_{\mathrm{f}}$ respectively. The initial value of the correlators is given by:
\begin{align}
& G_k(0) = v_{k,\mathrm{i}}^2,~~~~~~F_k(0) = u_{k,\mathrm{i}}v_{k,\mathrm{i}}, 
\end{align}
\noindent
where the pre-quench functions arising from the Bogolyubov transformation are defined as:
\begin{subequations}
\begin{align}
& u_{k,\mathrm{i}} = \cosh(\theta_{k,\mathrm{i}}); \\
& v_{k,\mathrm{i}} = \sinh(\theta_{k,\mathrm{i}}),\\
& 2\theta_{k,\mathrm{i}} = \mathrm{arctanh}\left(-\frac{B_{k,\mathrm{i}}}{A_{k,\mathrm{i}}}\right).
\end{align}
\end{subequations}
\noindent
On the top left panel of Fig.~\ref{srti}, the ETCCF $G_1(R,t)$ defined as $G_1(R,t) = \langle \hat{a}^{\dag}_R \hat{a}_0 \rangle_t - \langle \hat{a}^{\dag}_R \hat{a}_0 \rangle_0 = (1/L)\sum_k \cos(kR)[G_k(t)-G_k(0)]$ and deduced analytically using the EoM approach, while considering a bosonic reformulation of the TFI Hamiltonian $\hat{H}$, is plotted using the previous set of EoMs at Eq.~\eqref{EoMs_boson}. Using the HP transformation at Eq.~\eqref{hp}, the latter ETCCF is equivalent to the spin correlation function $G_{-+}(R,t) = \langle \hat{S}^{-}_R \hat{S}^{+}_0 \rangle_t - \langle \hat{S}^{-}_R \hat{S}^{+}_0 \rangle_0$. 

\paragraph{QP approach} We now move on to the QP approach based on the Heisenberg picture together with the bosonic Bogolyubov theory~\cite{despres_these,despres2018,cevolani2015,cevolani2016} . We first express the pre-quench and post-quench bosonic operators in reciprocal space denoted by $\hat{a}^{(\dag)}_{k,\mathrm{i}}$ and $\hat{a}^{(\dag)}_{k,\mathrm{f}}$ in terms of the pre-quench and post-quench (bosonic) Bogolyubov operators $\hat{b}^{(\dag)}_{k,\mathrm{i}}$ and $\hat{b}^{(\dag)}_{k,\mathrm{f}}$ respectively. Using the Bogolyubov transformation previously introduced, we have:
\begin{subequations}
\label{bogo_boson}
\begin{align}
& \hat{a}_{k,\mathrm{i}} = u_{k,\mathrm{i}} \hat{b}_{k,\mathrm{i}} + v_{k,\mathrm{i}} \hat{b}^{\dag}_{-k,\mathrm{i}}, \\
& \hat{a}_{k,\mathrm{f}}(t) = u_{k,\mathrm{f}} \hat{b}_{k,\mathrm{f}}(t) + v_{k,\mathrm{f}} \hat{b}^{\dag}_{-k,\mathrm{f}}(t). 
\end{align}
\end{subequations}
\noindent
Then, using the continuity of the time-evolved operators at $t=0$, we can express $\hat{b}^{(\dag)}_{k,\mathrm{f}}(0)$ in terms of $\hat{b}^{(\dag)}_{k,\mathrm{i}}$. It yields: 
\begin{subequations}
\begin{align}
\label{continuity_boson}
& \hat{b}_{k,\mathrm{f}}(0) = m_{k,\mathrm{i},\mathrm{f}} \hat{b}_{k,\mathrm{i}} + l_{k,\mathrm{i},\mathrm{f}} \hat{b}^{\dag}_{-k,\mathrm{i}};\\
& \hat{b}^{\dag}_{-k,\mathrm{f}}(0) = l_{k,\mathrm{i},\mathrm{f}} \hat{b}_{k,\mathrm{i}} + m_{k,\mathrm{i},\mathrm{f}} \hat{b}^{\dag}_{-k,\mathrm{i}}, 
\end{align}
\end{subequations}
\noindent
where the quasi-momentum-dependent functions $l$ and $m$ are defined as:
\begin{subequations}
\begin{align}
& m_{k,\mathrm{i},\mathrm{f}} = u_{k,\mathrm{i}}u_{k,\mathrm{f}} - v_{k,\mathrm{i}}v_{k,\mathrm{f}};\\
& l_{k,\mathrm{i},\mathrm{f}} = v_{k,\mathrm{i}}u_{k,\mathrm{f}} - v_{k,\mathrm{f}}u_{k,\mathrm{i}},
\end{align}
\end{subequations}
\noindent
Since $\hat{H}$ has a diagonalized form at Eq.~\eqref{diag_boson}, it immediately follows that: 
\begin{align}
\label{evolve_boson}
& \hat{b}_{k,\mathrm{f}}(t) = e^{-iE_{k,\mathrm{f}}t} \hat{b}_{k,\mathrm{f}}(0), ~~ \hat{b}^{\dag}_{-k,\mathrm{f}}(t) = e^{iE_{k,\mathrm{f}}t} \hat{b}^{\dag}_{-k,\mathrm{f}}(0),
\end{align}
\noindent
Finally, using Eqs.~\eqref{bogo_boson}, \eqref{evolve_boson} and \eqref{continuity_boson}, the correlator $G_k(t)$ has the analytical expression:
\begin{widetext}
\begin{align}
\label{corr_boson}
& G_k(t) =  (u_{k,\mathrm{f}}  l_{k,\mathrm{i},\mathrm{f}})^2 + (v_{k,\mathrm{f}} m_{k,\mathrm{i},\mathrm{f}})^2 + 2u_{k,\mathrm{f}}v_{k,\mathrm{f}}m_{k,\mathrm{i},\mathrm{f}}l_{k,\mathrm{i},\mathrm{f}}\cos(2E_{k,\mathrm{f}}t).
\end{align}
\end{widetext}
\noindent
From the latter, the ETCCF $G_{-+} = G_1$ takes the form: 
\begin{subequations}
\label{qp_boson}
\begin{align}
& G_1(R,t) = \frac{1}{L}\sum_k  F_{k} \cos(kR)\sin^2(E_{k,\mathrm{f}}t);\\
& F_{k} = -4u_{k,\mathrm{f}} v_{k,\mathrm{f}}m_{k,\mathrm{i},\mathrm{f}} l_{k,\mathrm{i},\mathrm{f}}
\end{align}
\end{subequations}
\noindent
which can be cast into the generic form already discussed. Indeed, using the thermodynamic limit, i.e. $L \rightarrow + \infty$, we get:
\begin{subequations}
\begin{align}
& G_1(R,t) \sim \int_{\mathcal{B}}\mathrm{d}k~ S^{(1)}_k \left[e^{i(kR+2E_{k,\mathrm{f}}t)} + e^{i(kR-2E_{k,\mathrm{f}}t)} \right],  \\
& S^{(1)}_{k} = u_{k,\mathrm{f}} v_{k,\mathrm{f}}m_{k,\mathrm{i},\mathrm{f}} l_{k,\mathrm{i},\mathrm{f}}
\end{align}
\end{subequations}
\noindent
On the bottom left panel of Fig.~\ref{srti}, the ETCCF $G_1(R,t)$ deduced analytically using the QP approach while considering bosonic Bogolyubov excitations is plotted using Eq.~\eqref{qp_boson}. We find a perfect benchmark of the space-time one-body correlation function $G_1(R,t)$ between the QP and EoMs theoretical approaches. Note that by using the same procedure than previously, i.e. the QP approach, the analytical expression of the second quadratic correlator $F_k(t)$ can be unraveled. Then, by taking the time derivative of the latter and the former, we have to recover the set of EoMs at Eq.~\eqref{EoMs_boson}. \\

\subsection{Fermionic reformulation}
\label{subsec:fermion}
We now turn to the case of the fermionic reformulation of the TFI Hamiltonian $\hat{H}$. When considering the paramagnetic, or equivalently the $z$-polarized phase, the low-lying excitations can be seen as
fermionic Bogolyubov ones. We first discuss its diagonalized form before investigating its quench dynamics properties using both the EoM and QP approaches. For the diagonalized form of $\hat{H}$ 
as well as the EoM approach, we will work along the lines of Ref.~\cite{despres2024} and present the main features.

\paragraph{Fermionic Bogolyubov theory} To unveil the quadratic Fermi form, we rely on the Jordan-Wigner (JW) transformation suitable for the paramagnetic phase given by: 
\begin{align}
& \hat{S}^+_R = \hat{K}_R \hat{c}_R,~~~~~~\hat{S}^{-}_R = \hat{K}_R \hat{c}^{\dag}_R,~~~~~~\hat{S}^z_R = \frac{1}{2} - \hat{c}^{\dag}_R \hat{c}_R,
\label{jw}
\end{align}
where $\hat{c}^{\dag}_R$ ($\hat{c}_R$) denotes the creation (annihilation) fermionic operator acting on the lattice site $R \in \mathbb{N}$. $\hat{K}_R = \prod_{R'=0}^{R-1}(1-2\hat{c}^{\dag}_{R'}\hat{c}_{R'})$ refers to the non-local string operator taking care of the anticommutativity.
The JW transformation, similarly to the HP transformation, requires
$\langle \hat{c}^{\dag}_R \hat{c}_R \rangle_t \ll 1$, $\forall (R,t) \in \mathbb{N} \times \mathbb{R}$. Again, the latter requirement is verified since a sudden global quench confined within the paramagnetic phase of the TFI chain is considered.
It yields the following quadratic Fermi form:
\begin{equation}
\hat{H} = \sum_{k}\mathfrak{A}_{k}\left(\hat{c}^{\dag}_{k} \hat{c}_{k}-\hat{c}_{-k}\hat{c}^{\dag}_{-k}\right)+\mathfrak{B}_{k} \left(\hat{c}^{\dag}_{k} \hat{c}^{\dag}_{-k}+\hat{c}_{k} \hat{c}_{-k} \right),
\label{eq_bhm_quadratic_fermion}
\end{equation}
\noindent
with $\mathfrak{A}_{k} = (h/2) + (J/4)\cos(k)$ and $\mathfrak{B}_{k} = -i(J/4)\sin(k)$. Note that the Hamiltonian $\hat{H}$ at Eq.~\eqref{eq_bhm_quadratic_fermion} can be rewritten using a fermionic Bogolyubov-de-Gennes Hamiltonian.
We then use the following Bogolyubov transformation while introducing fermionic Bogolyubov operators denoted by $\hat{f}^{(\dag)}$:
\begin{equation}
\hat{c}_k = \mathfrak{u}_k \hat{f}_{k} + i\mathfrak{v}_k \hat{f}^{\dag}_{-k},~~~~~~\hat{c}^{\dag}_{-k} = i\mathfrak{v}_k \hat{f}_{k} + \mathfrak{u}_k \hat{f}^{\dag}_{-k},
\label{tr_fermion}
\end{equation}
where $(\mathfrak{u}_k$, $\mathfrak{v}_k) \in \mathbb{R}^2$ with  $\mathfrak{u}_k$ and $\mathfrak{v}_k$ being an even and odd momentum-dependent function respectively. The latter are defined as: 
\begin{subequations}
\begin{align}
& \mathfrak{u}_k = \cos\left(\phi_k \right); \\
& \mathfrak{v}_k = \sin\left(\phi_k \right), \\
& 2\phi_k = \mathrm{arctan}\left(i\frac{\mathfrak{B}_k}{\mathfrak{A}_k}\right).
\end{align} 
\end{subequations}
\noindent
Injecting the latter permits to diagonalize $\hat{H}$ which reads as: 
\begin{equation}
\hat{H} = \sum_k \mathfrak{E}_k \hat{f}^{\dag}_k \hat{f}_k,~~\mathfrak{E}_k = 2\mathrm{sgn}(\mathfrak{A}_k)\sqrt{\mathfrak{A}_k^2 -\mathfrak{B}_k^2}. 
\label{diag_fermion}
\end{equation} 

\paragraph{EoM approach} We first consider the EoM approach, before moving on to the QP approach, while taking advantage of the fermionic reformulation. For both theoretical approaches, the procedure is similar to the one previously introduced where the bosonic reformulation was used. Regarding the EoM approach, we introduce the two relevant quadratic fermionic correlators $\mathfrak{G}$ and $\mathfrak{F}$ defined as $\mathfrak{G}_k(t) = \langle \hat{c}^{\dag}_k \hat{c}_k \rangle_t$
and $\mathfrak{F}_k(t) = \langle \hat{c}_k \hat{c}_{-k} \rangle_t$. Relying on the Heisenberg picture to deduce theoretically the time derivative of both correlators and injecting Eq.~\eqref{eq_bhm_quadratic_fermion} while using the fermionic Wick theorem together with the momentum conservation, we deduce the following set of EoMs: 
\begin{subequations}
\label{EoMs_fermion}
\begin{align}
\frac{\mathrm{d}}{\mathrm{d}t} \mathfrak{G}_k(t) =&~ 4i\mathfrak{B}_{k,\mathrm{f}}\operatorname{Re}[\mathfrak{F}_k(t)],\\
\frac{\mathrm{d}}{\mathrm{d}t} \mathfrak{F}_k(t) =&~ -4i\mathfrak{A}_{k,\mathrm{f}}\mathfrak{F}_k(t) + 2i\mathfrak{B}_{k,\mathrm{f}}[1-2\mathfrak{G}_k(t)].
\end{align}
\end{subequations}
\noindent
The latter EoM approach has been already used to study the quench dynamics of the non-Hermitian TFI chain confined within the $z$-polarized phase, see Ref.~\cite{despres2024}. In this paper, by fixing $\gamma=0$, we recover the EoMs at Eq.~\eqref{EoMs_fermion}.
Similarly as for the bosonic reformulation, the coefficients $\mathfrak{A}_{k,\alpha}$ and $\mathfrak{B}_{k,\alpha}$ where $\alpha \in \{\mathrm{i},\mathrm{f}\}$ denote the initial (index i) and the final (index f) momentum-dependent functions valid for the pre-quench and post-quench TFI Hamiltonians reformulated as quadratic fermionic ones. The latter will depend on $h_{\mathrm{i}}$ and $h_{\mathrm{f}}$ respectively. The initial value of each correlator is given by:
\begin{align}
& \mathfrak{G}_k(0) =  \mathfrak{v}_{k,\mathrm{i}}^2,~~~~~~\mathfrak{F}_k(0) = -i \mathfrak{u}_{k,\mathrm{i}} \mathfrak{v}_{k,\mathrm{i}}, 
\end{align}
\noindent
where the pre-quench functions $\mathfrak{u}_{k,\mathrm{i}}$ and $\mathfrak{v}_{k,\mathrm{i}}$ are defined as follows:
\begin{subequations}
\begin{align}
& \mathfrak{u}_{k,\mathrm{i}} = \cos(\phi_{k,\mathrm{i}}); \\
& \mathfrak{v}_{k,\mathrm{i}} = \sin(\phi_{k,\mathrm{i}}),\\
& 2\phi_{k,\mathrm{i}} = \mathrm{arctan}\left(i\frac{\mathfrak{B}_{k,\mathrm{i}}}{\mathfrak{A}_{k,\mathrm{i}}}\right).
\end{align}
\end{subequations}
\noindent
We now discuss the $G_{-+}(R,t) = g_{-+}(R,t) - g_{-+}(R,0)$ spin correlation function where $g_{-+}(R,t) = \langle \hat{S}^{-}_R \hat{S}^{+}_0 \rangle_t$. Using the JW transformation, the latter is given by: 
\begin{align}
g_{-+}(R,t) =& \langle \hat{K}_R \hat{c}^{\dag}_R \hat{K}_0 \hat{c}_0 \rangle_t; \nonumber \\
g_{-+}(R,t) =& \left \langle \left( \prod_{R'=0}^{R-1}1-2\hat{c}^{\dag}_{R'}\hat{c}_{R'} \right)\hat{c}^{\dag}_R \hat{c}_0 \right \rangle_t,
\end{align}
\noindent
using $ \hat{K}_0 = \mathbb{I}_{R = 0} \equiv 1$ by definition. Since the dynamics is confined within the $z$-polarized phase implying $\langle \hat{c}^{\dag}_R \hat{c}_R \rangle_t \ll 1$ and considering the expression of $g_{-+}$, we may approximate the string operator as $\hat{K}_R \simeq \mathbb{I}_R \equiv 1$. Finally and similarly to the bosonic case, the latter spin correlation function can be seen as the one-body correlation function, i.e. $g_{-+}(R,t) = g_1(R,t)$ where $g_1(R,t) = \langle \hat{c}^{\dag}_R \hat{c}_0 \rangle_t$. The (fermionic) equal-time connected one-body correlation function $G_1$ is given here by $G_1(R,t) = g_1(R,t) - g(R,0)$.  
On the top right panel of Fig.~\ref{srti}, the ETCCF $G_{-+}(R,t) = g_{-+}(R,t)  - g_{-+}(R,0)$, or equivalently $G_1(R,t)$ defined as $G_1(R,t) = (1/L)\sum_k \cos(kR)[\mathfrak{G}_k(t)-\mathfrak{G}_k(0)]$, deduced analytically using the EoM approach while considering a fermionic reformulation of the TFI Hamiltonian $\hat{H}$, is plotted using the previous set of EoMs at Eq.~\eqref{EoMs_fermion}. The latter is in very good agreement with both theoretical results of $G_{-+}$ deduced from the bosonic EoM and QP approaches.

\begin{figure*}[ht]
\centering
\includegraphics[scale = 0.445]{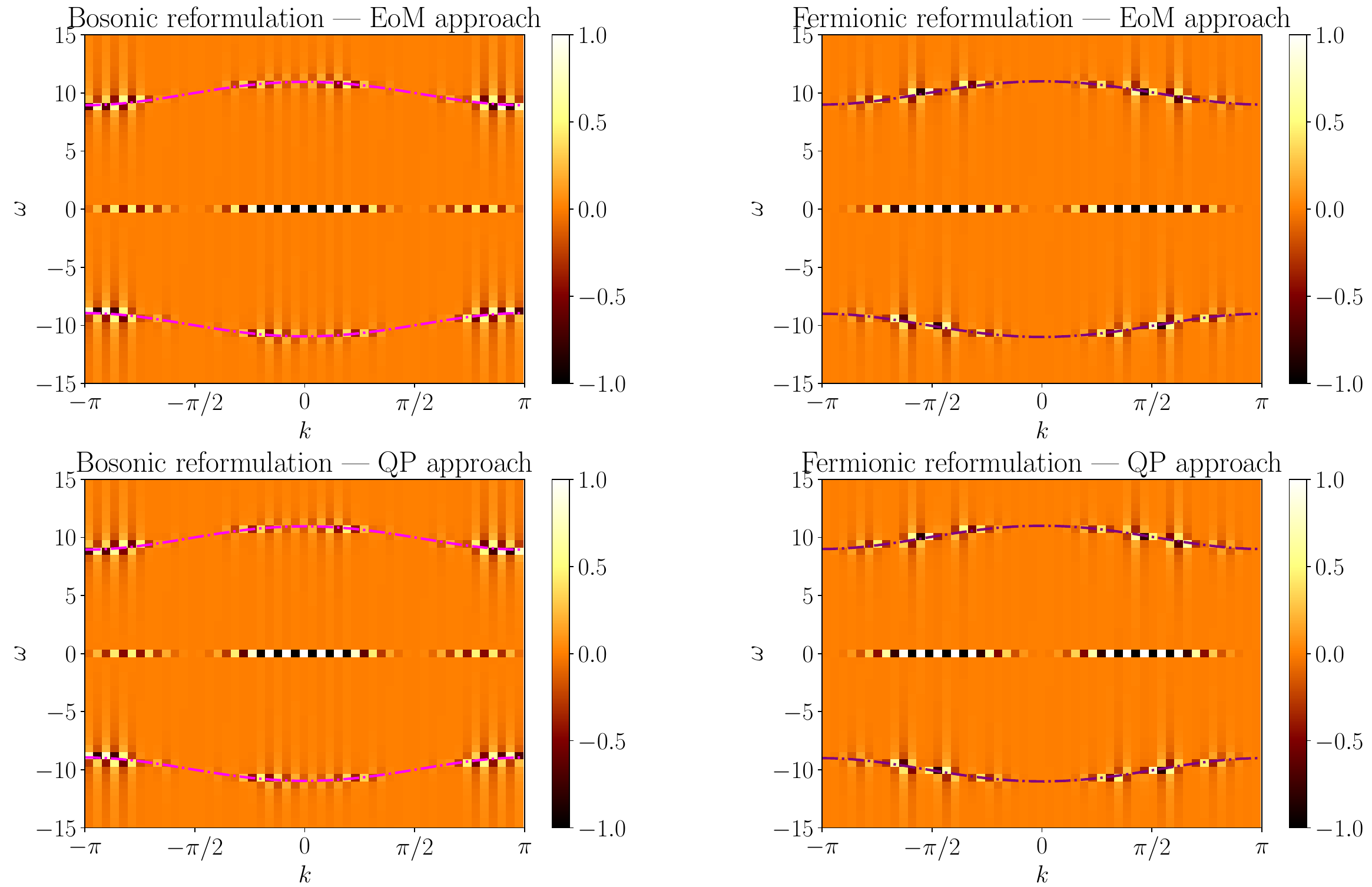}
\caption{Quench spectroscopy method applied to the TFI chain: a weak sudden global quench confined in the paramagnetic phase is considered for which the QSF  $S_k(\omega)$ associated to the one-body ETCCF $G_1(R,t)$ is investigated.
The latter is plotted as a function of the energy $\omega$ ($\hbar = 1$) and the momentum $k$ with $k \in \mathcal{B} = [-\pi,\pi]$ ($a = 1$) where $\mathcal{B}$ denotes the first Brillouin zone. For clarity, the space-time correlation pattern of $G_1$
for the calculation of $S_k(\omega)$ is plotted for $R \leq 0$ and $t \geq 0$ as well to recover the second half of the low-lying excitation spectrum. $S_k(\omega)$ is plotted here using the bosonic and fermionic reformulations of the 1D TFI Hamiltonian. (Top left panel) bosonic EoM approach, (bottom left panel) bosonic QP approach, (top right panel) fermionic EoM approach, (bottom right panel) fermionic QP approach. On the left panels, the dashed-dotted magenta lines correspond to twice the post-quench low-lying excitation spectrum of the TFI chain confined in the $z$-polarized phase deduced from the bosonic reformulation, i.e. $\pm 2 E_{k,\mathrm{f}}$ (see Eq.~\eqref{diag_boson}); whereas, on the right panels, the dashed-dotted purple lines correspond to twice the post-quench low-lying excitation spectrum of the TFI chain confined in the $z$-polarized phase deduced from the fermionic reformulation, i.e. $\pm 2 \mathfrak{E}_{k,\mathrm{f}}$ (see Eq.~\eqref{diag_fermion}). Note that both 
theoretical spectra are identical at first order in $J/h$. For all the different panels, the parameters are: $J = 1$, $h_{\mathrm{i}} = 8$, $h_{\mathrm{f}} = 5$ and a very good agreement is found between the four cases.}
\label{quench_spectro_srtfi}
\end{figure*}

\paragraph{QP approach} We now move on to the fermionic QP approach where the procedure is similar to the bosonic one. We first express the pre-quench and post-quench fermionic operators in reciprocal space denoted by $\hat{c}^{(\dag)}_{k,\mathrm{i}}$ and $\hat{c}^{(\dag)}_{k,\mathrm{f}}$ in terms of the pre-quench and post-quench (fermionic) Bogolyubov operators $\hat{f}^{(\dag)}_{k,\mathrm{i}}$ and $\hat{f}^{(\dag)}_{k,\mathrm{f}}$ respectively.
Using the Bogolyubov transformation at Eq.~\eqref{tr_fermion}, we have:
\begin{subequations}
\label{bogo_fermion}
\begin{align}
& \hat{c}_{k,\mathrm{i}} = \mathfrak{u}_{k,\mathrm{i}} \hat{f}_{k,\mathrm{i}} + i\mathfrak{v}_{k,\mathrm{i}} \hat{f}^{\dag}_{-k,\mathrm{i}}, \\
& \hat{c}_{k,\mathrm{f}}(t) = \mathfrak{u}_{k,\mathrm{f}} \hat{f}_{k,\mathrm{f}}(t) + i\mathfrak{v}_{k,\mathrm{f}} \hat{f}^{\dag}_{-k,\mathrm{f}}(t). 
\end{align}
\end{subequations}
\noindent
Then, using the continuity of the time-evolved fermionic operators at $t=0$, we can express $\hat{f}^{(\dag)}_{k,\mathrm{f}}(0)$ in terms of $\hat{f}^{(\dag)}_{k,\mathrm{i}}$. It yields: 
\begin{subequations}
\begin{align}
\label{continuity_fermion}
& \hat{f}_{k,\mathrm{f}}(0) = \mathfrak{m}_{k,\mathrm{i},\mathrm{f}} \hat{f}_{k,\mathrm{i}} + \mathfrak{l}_{k,\mathrm{i},\mathrm{f}} \hat{f}^{\dag}_{-k,\mathrm{i}};\\
& \hat{f}^{\dag}_{-k,\mathrm{f}}(0) = \mathfrak{l}_{k,\mathrm{i},\mathrm{f}} \hat{f}_{k,\mathrm{i}} + \mathfrak{m}_{k,\mathrm{i},\mathrm{f}} \hat{f}^{\dag}_{-k,\mathrm{i}}, 
\end{align}
\end{subequations}
\noindent
where the quasi-momentum-dependent functions $\mathfrak{l}$ and $\mathfrak{m}$ are defined as:
\begin{subequations}
\begin{align}
& \mathfrak{m}_{k,\mathrm{i},\mathrm{f}} = \mathfrak{u}_{k,\mathrm{i}}\mathfrak{u}_{k,\mathrm{f}} + \mathfrak{v}_{k,\mathrm{i}}\mathfrak{v}_{k,\mathrm{f}};\\
& \mathfrak{l}_{k,\mathrm{i},\mathrm{f}} = i(\mathfrak{v}_{k,\mathrm{i}}\mathfrak{u}_{k,\mathrm{f}} - \mathfrak{v}_{k,\mathrm{f}}\mathfrak{u}_{k,\mathrm{i}}),
\end{align}
\end{subequations}
\noindent
Since $\hat{H}$ has a diagonalized form at Eq.~\eqref{diag_fermion}, it immediately follows that: 
\begin{align}
\label{evolve_fermion}
& \hat{f}_{k,\mathrm{f}}(t) = e^{-i\mathfrak{E}_{k,\mathrm{f}}t} \hat{f}_{k,\mathrm{f}}(0), ~~ \hat{f}^{\dag}_{-k,\mathrm{f}}(t) = e^{i\mathfrak{E}_{k,\mathrm{f}}t} \hat{f}^{\dag}_{-k,\mathrm{f}}(0),
\end{align}
\noindent
Finally, using Eqs.~\eqref{bogo_fermion}, \eqref{evolve_fermion} and \eqref{continuity_fermion}, the correlator $\mathfrak{G}_k(t)$ has the following analytical expression:
\begin{widetext}
\begin{align}
\label{corr_fermion}
& \mathfrak{G}_k(t) =  - (\mathfrak{u}_{k,\mathrm{f}} \mathfrak{l}_{k,\mathrm{i},\mathrm{f}})^2 + (\mathfrak{v}_{k,\mathrm{f}}  \mathfrak{m}_{k,\mathrm{i},\mathrm{f}})^2  - 2i\mathfrak{u}_{k,\mathrm{f}}\mathfrak{v}_{k,\mathrm{f}}\mathfrak{m}_{k,\mathrm{i},\mathrm{f}}\mathfrak{l}_{k,\mathrm{i},\mathrm{f}}\cos(2\mathfrak{E}_{k,\mathrm{f}}t).
\end{align}
\end{widetext}
\noindent
From the latter, the ETCCF $G_{-+} = G_1$ takes the form: 
\begin{subequations}
\label{qp_fermion}
\begin{align}
& G_1(R,t) = \frac{1}{L}\sum_k \mathfrak{F}_{k} \cos(kR)\sin^2(\mathfrak{E}_{k,\mathrm{f}}t);\\
& \mathfrak{F}_{k} = 4i\mathfrak{u}_{k,\mathrm{f}} \mathfrak{v}_{k,\mathrm{f}}\mathfrak{m}_{k,\mathrm{i},\mathrm{f}} \mathfrak{l}_{k,\mathrm{i},\mathrm{f}}
\end{align}
\end{subequations}
\noindent
which can be cast into the generic form using the thermodynamic limit, i.e. $L \rightarrow + \infty$. We find:
\begin{subequations}
\begin{align}
& G_1(R,t) \sim \int_{\mathcal{B}} \mathrm{d}k~ \mathfrak{S}^{(1)}_k \left[e^{i(kR+2\mathfrak{E}_{k,\mathrm{f}}t)} + e^{i(kR-2\mathfrak{E}_{k,\mathrm{f}}t)} \right],  \\
& \mathfrak{S}^{(1)}_{k} = -i\mathfrak{u}_{k,\mathrm{f}} \mathfrak{v}_{k,\mathrm{f}}\mathfrak{m}_{k,\mathrm{i},\mathrm{f}} \mathfrak{l}_{k,\mathrm{i},\mathrm{f}}
\end{align}
\end{subequations}
\noindent
On the bottom right panel of Fig.~\ref{srti}, the ETCCF $G_1(R,t)$ deduced analytically using the QP approach while considering low-lying fermionic Bogolyubov excitations is plotted using Eq.~\eqref{qp_boson}. We find a perfect benchmark of the space-time one-body correlation function $G_1(R,t)$ between the QP and EoMs theoretical approaches. Note that by using the same scheme than previously, i.e. the QP approach, the analytical expression of the second quadratic
correlator $\mathfrak{F}_k(t)$ can be unraveled. Then, by taking the time derivative of the latter and the former correlators, we should recover the set of EoMs at Eq.~\eqref{EoMs_boson}. As expected from the generic form of the ETCCFs for weak sudden global quenches in short-range interacting quantum lattice models, we have found a twofold ballistic correlation spreading: a linear CE as well as a linear series of local extrema. Consequently, the quench spectroscopy method should apply as previously discussed in the case of the Bose-Hubbard chain confined in the SF phase. \\

To verify the previous statement, on Fig.~\ref{quench_spectro_srtfi}, the QSF $S_k(\omega)$ associated to the ETCCF $G_1(R,t)$ previously investigated (see Fig.~\ref{srti}) is plotted while using the bosonic and fermionic reformulations and the EoM and QP theoretical approaches. A very good agreement is found between all the different cases. As expected, we recover twice the low-lying excitation spectrum of the post-quench Hamiltonian, i.e. $2E_{k,\mathrm{f}}$ (or equivalently $2\mathfrak{E}_{k,\mathrm{f}}$). 

To summarize, we have shown here, with the case study of the TFI chain confined in the paramagnetic phase and having a bosonic and fermionic reformulations, a benchmark of the quench dynamics properties between both reformulations and between both theoretical methods namely the EoM and QP approaches. This permits to show, on one hand, the robustness of our analytical results and on the other hand, the universality of the quench dynamics features of isolated Hermitian quantum
lattice models. More precisely, for any quantum system displaying a generic Bose or Fermi form, i.e. its Hamiltonian can be reformulated using a bosonic or fermionic Bogolyubov-de-Gennes Hamiltonian, this leads to a twofold ballistic, i.e. linear, structure of the space-time correlations. The latter is a sufficient but not necessary condition to apply the quench spectroscopy method in order to unveil the low-lying excitation spectrum of the post-quench Hamiltonian.   

\section{Conclusion}
\label{sec:conclusions}
In this work we have discussed the quench spectroscopy method for dissipative, isolated non-Hermitian and Hermitian quantum lattice models theoretically. For dissipative and isolated non-Hermitian quantum systems, we considered the case study of the open Bose-Hubbard chain confined in the superfluid phase where the dissipation is characterized by on-site two-body losses and the non-Hermitian transverse-field Ising chain confined in the paramagnetic phase respectively. For isolated Hermitian quantum lattice models, we used the Hermitian version of the latter quantum system.  

Regarding the dissipative Bose-Hubbard chain, we have first introduced the theoretical method based on the equation-of-motion approach. We then benchmarked the latter with the quasiparticle theory for two distinct equal-time connected correlation functions while considering the isolated Bose-Hubbard chain. The latter functions are $G_1$ the one-body correlation function as well as $G_2$ the density-density correlation function giving insight on the phase and density fluctuations of the quantum system respectively. We then investigated the double quench dynamics of the Bose-Hubbard chain where both the repulsive interactions and the dissipations are quenched. We finally studied the quench spectroscopy of this quantum model by calculating the quench spectral function associated to the one-body and density-density correlation functions. 

Concerning the non-Hermitian transverse-field Ising chain confined in the paramagnetic phase, its quench dynamics has been investigated using the equation-of-motion approach and by considering an equal-time spin-spin correlation function in order
to apply the quench spectroscopy method. 

For the Hermitian case, we considered the Hermitian version of the previous model while considering the same gapped quantum phase. The advantage of this quantum model is twofold: the possibility to express the Hamiltonian in a quadratic fermionic and bosonic form and the possibility to rely on the equation-of-motion and quasiparticle theoretical approaches. This has permitted to unveil that for generic isolated and Hermitian quantum lattice models displaying a quadratic form, the quench spectroscopy method is suitable and reliable. To summarize, in the framework of weak sudden global quantum quenches, this paper paves the way to the possibility of applying the quench spectroscopy method to dissipative, isolated non-Hermitian and isolated Hermitian quadratic quantum lattice models. 

As an extension of the presented research work, the generalizability of the quench spectroscopy method is an interesting research topic. The latter is likely to be extended to dissipative fermionic lattice models, dissipative spin lattice models and to 
their reformulation in the continuum and for a higher dimensionality of the lattice or the space respectively. This new spectroscopy method is also expected to be reliable for dissipative quantum systems involving long-range interactions and for other Lindblad jump operators, i.e. other kind of loss processes. 

While considering large timescales, another possible extension would be to investigate the applicability of the quench spectroscopy method while using quantum computing platforms. In a first time, this can be done at the architecture level using a digital twin of the quantum computer. In a second time, provided that the latter investigation is conclusive, one may consider to study this topic at the experimental level. Since the quench spectroscopy method requires a relatively large quantum system, i.e. many logical qubits, suitable experimental platforms are neutral-atom-based quantum computers and trapped ion quantum computers; the latter being derived from well-known quantum simulators.

\section*{Acknowledgments}
The Author acknowledges funding from the European Research Council (ERC) under the European Union's Horizon 2020 research and innovation programme (Grant agreement No. 101002955 -- CONQUER),  from the Région Île-de-France in the framework of DIM QuanTiP and from the ANR project LOQUST ANR-23-CE47-0006-02. The Author wishes to acknowledge M. Cheneau for fruitful discussions concerning the experimental feasibility of the quench spectroscopy method using cold-atom-based quantum simulators. The Author is also grateful to L. Sanchez-Palencia for useful comments regarding the manuscript. Finally, the Author would like to thank M. Schirò and L. Mazza for useful discussions regarding the dissipative quench dynamics of the Bose-Hubbard model. 

\appendix
\section{The non-connected equal-time density-density correlation function in the SF-mean-field regime and the condensate density in the non-interacting case}
\label{g2}
In what follows, we discuss the analytical expression of $g_2(R,t)$ and $g_1(R,t)$ the non-connected density-density and non-connected one-body correlation functions. As previously mentioned, the latter can be found using the set of EoMs defined at Eq.~\eqref{EoMs}. As a reminder, $n(t) = N(t)/L = \langle \hat{N} \rangle_t/L$ refers to the time-dependent density (or equivalently the time-dependent filling) with $\hat{N} = \sum_R \hat{n}_R$ denoting the total occupation number operator; $g_2(R,t) = \langle \hat{n}_R \hat{n}_0 \rangle_t$ and $g_1(R,t) = \langle \hat{b}^{\dag}_R \hat{b}_0 \rangle_t$. We remind the reader that the latter correlation functions respectively permit to characterize the density and phase fluctuations and can be measured in experimental setups based on ultra-cold atoms using fluorescence microscopy imaging and time-of-flight techniques~\cite{cheneau2012,trotzky2012,langen2013,geiger2014}. To derive their analytical expression, a same theoretical approach is performed namely: (i) we first express them in the reciprocal space (ii) we then consider a decoupling of the condensate mode $k = 0$ from the finite modes $k \neq 0$ (iii) we simplify the expression using the standard mean-field approximation, i.e. $N(t) \gg \sum_{k \neq 0} G_k(t)$, as well as the product state approximation in momentum space $\langle \hat{n}_0^2 \rangle_t = \langle \hat{n}_0 \rangle_t^2$. The two previous approximations are valid provided that the SF-mean-field regime is considered as well as small observation times. This mathematical procedure leads to:
\begin{widetext}
\begin{subequations}
\label{eq_g1_g2}
\begin{align}
g_2(R,t) =&~ \frac{1}{L^2} \Big \{ G_0(t)^2 + \sum_{k\neq0} \left(e^{ikR} F_0(t)F_k(t)^* + \mathrm{h.c} \right) + e^{-ikR}G_0(t) + e^{ikR}G_k(t) + 2[1+\cos(kR)]G_0(t)G_k(t) \Big \},\\
g_1(R,t) =&~ \frac{1}{L} \Big[G_0(t) + \sum_{k\neq0}\cos(kR)G_k(t) \Big].
\end{align}
\end{subequations}
\end{widetext}

\noindent
Note that this theoretical expression for $g_1$ is consistent with our previous findings, see Eq.~\eqref{G1_isolated} and the discussion beforehand. In what follows, we explain how to simplify the previous analytical form of $g_2$.
To do so, we rely on the following approximations: $\operatorname{Re}(F_0(t)) \gg |\operatorname{Im}(F_0(t))|$, $|\sum_{k\neq0}\operatorname{Re}(F_k(t))| \gg |\sum_{k \neq 0} \operatorname{Im}(F_k(t))|$ and $N(t) \gg \sum_{k \neq 0} G_k(t)$ \cite{despresnew}. The latter are deduced from the EoMs at Eq.~\eqref{EoMs} and valid in the SF-mean-field regime while considering the thermodynamic limit, i.e. $L\rightarrow +\infty$, and small observations times.
This finally leads to Eq.~\eqref{theory_g2}. Note also that the condensate density for the non-interacting case, i.e. $U = 0$, given at Ref.~\cite{despresnew} can be recovered using the previous expression of $g_1$ and $g_2$. 
Indeed, for $U = 0$, we have $N(t) = G_0(t)$ and using the standard mean-field approximation, i.e. $N(t) \gg \sum_{k \neq 0}G_k(t)$, we can significantly simplify the expression of $g_1$ and $g_2$ given at Eq.~\eqref{eq_g1_g2}. It yields: 
\begin{subequations}
\label{simplified_g1_g2}
\begin{align}
& g_2(R,t) = n(t)\Big[n(t) + \frac{1}{L} \sum_{k \neq 0} e^{-ikR} \Big], \\
& g_1(R,t) = n(t).
\end{align}
\end{subequations}

\noindent
In the thermodynamic limit, i.e. $L \rightarrow + \infty$, and since $R \in \mathbb{N}$, we have $(1/L)\sum_{k \neq 0} e^{-ikR} = \delta_{R,0}$ and thus $g_2$ takes the simple form $g_2(R,t) = n(t)[n(t) + \delta_{R,0}]$. By summing over the momentum $k$ the EoM associated to $G_k(t) = \langle \hat{n}_k \rangle_t$ the time-dependent occupation number in reciprocal space at Eq.~\eqref{lindblad_nl}, while considering the 1D case as well as on-site two-body losses, the EoM associated to the time-dependent density $n(t)$ the total occupation number can be deduced~\cite{despresnew} and it yields:  
\begin{align}
& \frac{\mathrm{d}}{\mathrm{d}t}n(t) = -\frac{2 \gamma}{L} \sum_R \langle \hat{b}^{\dag}_R \hat{b}^{\dag}_R \hat{b}_R \hat{b}_R \rangle_t.
\end{align}
\noindent
Starting from the latter expression and using the canonical commutation rules as well as the translational invariance of the BH model together with Eq.~\eqref{simplified_g1_g2}, we obtain:
\begin{align}
& \frac{\mathrm{d}}{\mathrm{d}t}n(t) = -2\gamma \left[g_2(0,t) - g_1(0,t) \right] = -2\gamma n(t)^2,
\end{align}

\noindent
which can be solved analytically and we finally recover the result from Ref.~\cite{despresnew} regarding the condensate density $n(t)$ for the non-interacting case, i.e. $U = 0$, namely:
\begin{equation}
n(t) = \frac{\bar{n}}{1+2\gamma \bar{n}t},
\end{equation}

\noindent
where $\bar{n} = \bar{n}(0)$ refers to the initial filling of the lattice chain. We stress that the previous theoretical form for the condensate density $n(t)$ is valid provided that the thermodynamic limit as well as small observation times are considered. Note that a more generic analytical expression for the latter quantity is provided at Ref.~\cite{despresnew}. 

\bibliographystyle{revtex}
\bibliography{biblioJD}

\end{document}